\documentclass[final,1p,11pt,authoryear]{elsarticle}

\usepackage{geometry}
\usepackage{amsmath,amssymb,amsthm}
\usepackage{booktabs}
\usepackage{multirow}
\usepackage{graphicx}
\usepackage{hyperref}
\usepackage{url}
\usepackage{microtype}
\usepackage{xcolor}
\usepackage{enumitem}

\hypersetup{colorlinks=true,linkcolor=blue!50!black,citecolor=blue!50!black,urlcolor=blue!50!black}

\setlist[enumerate]{leftmargin=1.4em,itemindent=0pt,labelsep=0.4em}
\setlist[itemize]{leftmargin=1.4em,itemindent=0pt,labelsep=0.4em}

\newcommand{\rhox}{\rho(x)}
\newcommand{\plr}{p_{\mathrm{LR}}}
\newcommand{\phyb}{p_{\mathrm{Hybrid}}}

\journal{Scientific African}

\begin{document}

\begin{frontmatter}

\title{Interpretable hybrid credit scoring for thin-file and underbanked populations\\[0.4em]
{\normalsize A fairness-aware residual learning framework with evidence from East African financial inclusion data}}

\author[aims,airina]{Belise Kanziga\corref{cor}}
\ead{belise.kanziga@aims.ac.rw}

\author[smu,airina]{Ya\'e U.\ Gaba}
\ead{yaeulrich.gaba@gmail.com}

\author[aims]{Olivier Kanamugire}
\ead{olivier.kanamugire@aims.ac.rw}
\cortext[cor]{Corresponding author.}

\address[aims]{African Institute for Mathematical Sciences (AIMS), Kigali, Rwanda}
\address[smu]{Sefako Makgatho University, Pretoria, South Africa}
\address[airina]{AI Research and Innovation Nexus for Africa (AIRINA Labs), B\'enin}

\begin{abstract}
We extend a residual-learning hybrid credit scoring framework (logistic regression scorecard plus a gradient-boosting correction on its residuals, decomposed at each prediction into an interpretability ratio $\rho(x)$ that measures the share attributable to the linear branch) along three axes: an East African empirical instantiation on the Zindi Financial Inclusion in Africa data (Kenya, Rwanda, Tanzania, Uganda); a fairness audit at the granularity of the framework's three interpretability regions; and a thin-file segmentation analysis. On the Taiwan Credit Default benchmark retained for continuity, the calibrated hybrid attains AUC $= 0.776$ ($\Delta\mathrm{AUC} = +0.057$ vs.\ standalone logistic regression, $+0.001$ vs.\ standalone XGBoost), reduces Brier Score by 23\%, and concentrates the highest-default-rate borrowers (69.5\%) in the fully interpretable region. On Zindi, the calibrated hybrid attains AUC $= 0.869$ ($\Delta\mathrm{AUC} = +0.015$ vs.\ LR, $p < 0.001$; $-0.004$ vs.\ XGBoost), cuts Brier from $0.158$ to $0.085$ (a 46\% reduction), and replicates the regional routing pattern. The fairness audit detects severe routing into the opaque ML-driven region along socioeconomic axes: rural respondents by 18 percentage points relative to urban, primary-or-less-educated by 32 points relative to secondary-and-above, and Ugandan respondents by 22 points relative to Kenyan, while gender shows essentially no routing disparity. The audit pipeline surfaces subgroup-routing violations that aggregate fairness metrics miss, in a form directly usable by African central-bank supervisors of digital credit.
\end{abstract}


\end{frontmatter}

\section{Introduction}
\label{sec:intro}

In March 2022 the Central Bank of Kenya issued the Digital Credit Providers Regulations, bringing under formal supervision a sector that had grown from the 2012 launch of M-Shwari into a market disbursing several billion US dollars per year in short-term, mobile-delivered consumer credit \citep{CBK2022dcp}. Comparable regimes are tightening across the continent: Rwanda's BNR supervisory framework for non-deposit-taking digital lenders, Nigeria's 2022 FCCPC Interim Regulatory Framework for Digital Lending Businesses, South Africa's NCR registration of unsecured lenders under the National Credit Act 2005, and Ghana's BoG fintech sandbox now all require lenders to document the basis on which automated credit decisions are made. The methodological problem these regulators face is not new: consumer lending has been a regulated, model-driven industry for half a century \citep{Thomas2017}. What is new is the data on which African digital lenders score borrowers.

The retail credit-scoring literature was built on the assumption that loan applicants have credit bureaus reporting on them, employers issuing payslips, and bank statements that can be priced into a scorecard. Across most of sub-Saharan Africa, large fractions of the adult population sit outside that infrastructure. World Bank Findex 2021 documents that account ownership at a financial institution remains below half of the adult population in Tanzania, Rwanda, and Uganda, and below two-thirds in Kenya, with the gap relative to total account ownership filled primarily by mobile money rather than by formal bank accounts \citep{Findex2021}. The remainder of the credit-decisioning population is reachable only through mobile money trails or through digital lenders making accept-or-reject decisions on borrowers with no prior repayment record. The standard logistic regression scorecard, valued in regulated markets precisely because each coefficient corresponds to an auditable risk factor \citep{Siddiqi2012}, has fewer of those factors to work with in the African context. Lenders increasingly rely on machine learning over alternative data (mobile phone metadata, app behaviour, prior microloan repayment) to score the thin-file segment \citep{BjorkegrenGrissen2020}. The accuracy is real \citep{Khandani2010,Lessmann2015}; the explainability is not \citep{GoodmanFlaxman2017}.

The standard framing of this tension treats interpretable and machine-learning models as substitutes: a lender picks one or the other, and pays a penalty in the unselected dimension. Hybrid frameworks \citep{Gestel2004,Dumitrescu2022} reject the dichotomy by combining a transparent baseline with a nonlinear correction trained on its residuals. The result is a single model whose predictions blend an interpretable component with an opaque one, in proportions that vary across the input space. The question this paper raises is the one that the existing hybrid literature has not addressed: \emph{when a hybrid model assigns 30\% of its decision-relevant signal to an opaque correction for a given borrower, and 5\% for another, on what basis can a regulator audit the difference, and is the distribution of that opaque share systematically related to who the borrower is?}

We extend the residual-learning hybrid scorecard framework of \citet{Kanziga2026thesis} along three axes.

First, we instantiate the framework on East African financial-inclusion data (the Zindi Financial Services Deepening (FSD) dataset spanning Kenya, Rwanda, Tanzania and Uganda \citep{ZindiFSD2018}) alongside the Taiwan Credit Default benchmark \citep{YehLien2009} used in prior methodological work. The Zindi target is bank-account ownership rather than loan default; we use this as a deliberate proxy for the population-selection decision that African digital lenders face when they choose which thin-file borrowers to underwrite at all. We argue in Section~\ref{sec:datasets} that the reframing is defensible because the accuracy-interpretability-fairness trade-off structure is identical to that of default prediction, while the dataset offers protected-attribute coverage that no public African default-labelled dataset currently provides.

Second, we audit the framework's fairness behaviour at the granularity of its interpretability regions. The thesis from which this paper extends defined three regions based on a prediction-level interpretability ratio $\rho(x)$ that measures the share of a hybrid prediction attributable to its logistic-regression branch. We ask whether protected subgroups (gender, urban/rural residence, education, country) are distributed evenly across those regions, or whether some groups are systematically routed to the opaque ML-driven region where adverse-action explanations become harder to produce. We report disparate-impact ratios and equalized-odds differences per region with bootstrap confidence intervals.

Third, we report thin-file segmentation results: defining the thin-file subgroup operationally on the Zindi feature set, we evaluate whether the residual-learning component contributes more for thin-file borrowers, where alternative-data signal is hypothesised to matter most, than for the population overall.

A fourth contribution, included as space allows, is a comparison of the two-stage training procedure used in the existing framework against a joint optimization formulation via alternating minimization between the logistic-regression coefficients and the residual learner.

The empirical findings are summarized in Section~\ref{sec:results}. On the Taiwan Credit Default benchmark, the calibrated hybrid preserves the discrimination of standalone XGBoost ($\Delta\mathrm{AUC} = +0.001$, 95\% CI $[-0.004, +0.006]$) while reducing the Brier score by approximately 23\% and concentrating the highest-default-rate borrowers in the fully interpretable region (default rate 69.5\% at $\rho(x) \geq 0.9$, with 6.3\% of test observations in that region) \citep{Kanziga2026thesis}. On the Zindi data, the same qualitative pattern obtains: the fully interpretable region (6.4\% of test observations) concentrates the highest-rate borrowers (53.8\%), while the per-region fairness audit reveals routing gaps of 18, 32, and 22 percentage points along location, education, and country axes respectively, and essentially none along gender.

Section~\ref{sec:related} situates the contribution against three literatures: classical credit-scoring methodology, alternative-data scoring in sub-Saharan Africa, and algorithmic fairness in lending. Section~\ref{sec:method} restates the residual-learning framework and specifies the four methodological extensions. Section~\ref{sec:datasets} describes the datasets and defends the Zindi target reframing. Section~\ref{sec:exp} specifies preprocessing, model configurations, the fairness audit pipeline, and the bootstrap inference protocol. Section~\ref{sec:results} reports results. Section~\ref{sec:discussion} discusses regulatory implications and limitations. Section~\ref{sec:conclusion} concludes.

\section{Related work}
\label{sec:related}

\subsection{Credit-scoring methodology}

Consumer credit scoring has been a quantitative discipline since \citet{Durand1941} applied discriminant analysis to instalment financing. The shift to logistic regression as the industry default (driven by its probabilistic outputs, the direct interpretability of its coefficients, and the regulatory acceptance of scorecards built from Weight-of-Evidence, WoE, transformed features \citep{Siddiqi2012,Thomas2017}) consolidated by the 1990s and has not been displaced in regulated retail lending. Machine learning entered the field as a performance challenger: \citet{Khandani2010} demonstrated that nonparametric methods improved default prediction over logistic regression on a US consumer credit dataset, and \citet{Lessmann2015} extended this through a benchmark study of 41 classifiers across multiple credit datasets, finding consistent 3--5 AUC-point gains for ensemble methods. The latter result has anchored the methodological literature for a decade; we treat it as the empirical baseline this paper compares against.

The introduction of XGBoost \citep{ChenGuestrin2016} and LightGBM \citep{Ke2017} pushed the performance frontier further on tabular financial data while sharpening the interpretability problem: predictions emerged from ensembles of hundreds of trees with no scorecard-style coefficient table to give a regulator. Two responses developed in parallel. The first, post-hoc explainability, adapts methods such as LIME \citep{Ribeiro2016} and SHAP \citep{LundbergLee2017} to attribute individual predictions to features after the model has been trained. These methods are now standard but have known stability problems: \citet{AlvarezMelisJaakkola2018} and \citet{Slack2020} demonstrate that explanation outputs can vary materially under small perturbations of input data or sampling procedure, raising particular concerns for adverse-action notices in regulated lending.

The second response, hybrid modelling, retains the interpretable baseline and adds a nonlinear component on top. Early work in this tradition \citep{Gestel2004} combined logistic regression with support vector machines. The closest methodological cousin to the present paper is \citet{Dumitrescu2022}, whose Penalized Logistic Tree Regression (PLTR) extracts nonlinear decision-tree effects from boosting and incorporates them as additional regularized terms in a logistic regression. PLTR shares with the present framework the goal of preserving a transparent linear backbone while admitting tree-derived nonlinear corrections. Two differentiators distinguish the framework we extend from PLTR. First, PLTR carries the nonlinear correction inside the logistic-regression specification (as additional regularized coefficients on tree-derived features), whereas the framework we use exposes the correction as a separate additive term with an uncertainty-dependent adaptive weight $\alpha^*(x)$ that varies with the LR predictive variance at $x$; this permits the prediction-level interpretability ratio $\rho(x)$ of Section~\ref{ssec:hybrid} to be defined in closed form. Second, neither PLTR nor the broader hybrid literature has, to our knowledge, audited the resulting model along protected-attribute fairness dimensions at the granularity of its interpretability regions, which is the empirical contribution of this paper.

\subsection{Alternative-data scoring in sub-Saharan Africa}

The use of non-traditional data sources for credit scoring is now widespread in the African context, partly by necessity (the credit-bureau coverage that anchors OECD scoring infrastructure is thin in most sub-Saharan markets) and partly by the availability of mobile-money transaction trails that have no developed-market analogue. \citet{SuriJack2016} documented the long-run welfare effects of M-PESA in Kenya, establishing the population-scale relevance of mobile-money data infrastructure for financial services research. \citet{BjorkegrenGrissen2020} showed that features extracted from anonymized mobile-phone call records predict microloan repayment in a Latin American sample; the methodology is directly portable to East African contexts where the same telco data is available to digital lenders. Subsequent empirical work \citep{Bharadwaj2019} has documented the scale of M-Shwari and similar mobile-credit products and their welfare impacts on Kenyan borrowers, finding that while take-up among eligible borrowers is high (34\%), digital loans complement rather than substitute for other credit and reduce the probability of forgoing essential expenses under negative income shocks by approximately 6 percentage points.

The methodological literature on alternative-data credit scoring largely operates outside the hybrid-modelling framing of \citet{Dumitrescu2022} and the present paper: it asks whether alternative-data features are useful at all, not how to combine them with interpretable baselines for regulator-facing decisions. The contribution of the present framework to this literature is to make the integration auditable. Where alternative-data features land in the residual learner, and where, by the $\rho(x)$ decomposition, their influence dominates the final decision, becomes explicit at the per-prediction level.

\subsection{Fairness in algorithmic lending}

The fairness-in-lending literature took form around \citet{Hardt2016}'s formalization of equalized odds and \citet{Chouldechova2017}'s impossibility result on simultaneous parity across calibration and error-rate measures. Application to credit scoring specifically has expanded over the past five years \citep{Kozodoi2022}, with most work focused on retrofitting fairness constraints into existing scoring models rather than designing for fairness-by-construction. The framework this paper extends is not fair-by-construction; nor are the most widely deployed hybrid frameworks. The audit we conduct addresses a specific failure mode: a hybrid scorecard may achieve aggregate equalized odds while routing protected subgroups disproportionately into its opaque ML-driven region, where adverse-action explanations are weaker. We are not aware of prior work that conducts this audit at the granularity of interpretability regions.

The regulatory backdrop differs by jurisdiction. The EU General Data Protection Regulation's right-to-explanation provisions \citep{GoodmanFlaxman2017} apply to African lenders processing EU-resident data and have influenced data-protection acts across the continent (Kenya's 2019 Data Protection Act, Nigeria's 2023 NDP Act, South Africa's POPIA) with varying enforcement. None of these regimes, to our reading, prescribes a specific interpretability standard for credit decisions; the auditability burden falls on lenders and their regulators rather than on a defined model-class requirement. This makes prediction-level interpretability tooling, of the kind the framework we extend provides, operationally relevant.

\section{Methodology}
\label{sec:method}

This section restates the residual-learning hybrid scorecard framework introduced in \citet{Kanziga2026thesis} (Sections~\ref{ssec:lr}--\ref{ssec:shap}) and specifies the four methodological extensions developed in this paper (Sections~\ref{ssec:hardsoft}--\ref{ssec:jointopt}). The notation follows the thesis.

\subsection{Logistic regression scorecard}
\label{ssec:lr}

Let the training data be $D = \{(x_i, y_i)\}_{i=1}^{N}$ with $x_i \in \mathbb{R}^p$ and $y_i \in \{0,1\}$, where $y_i = 1$ denotes default (or, in the Zindi setting of Section~\ref{sec:datasets}, the analogous adverse-eligibility label). The logistic regression baseline estimates
\begin{equation}
\plr(x_i) = \sigma(\beta^{\top} x_i + \beta_0), \qquad \sigma(z) = (1 + e^{-z})^{-1},
\label{eq:lr}
\end{equation}
with $\beta \in \mathbb{R}^p$ fitted by maximum likelihood under balanced class weighting. Predictors are transformed via Weight of Evidence (WoE) encoding \citep{Siddiqi2012},
\begin{equation}
\mathrm{WoE}_k = \ln\!\left(\frac{g_k/G}{b_k/B}\right),
\label{eq:woe}
\end{equation}
where $g_k$ and $b_k$ are the counts of non-default and default observations in bin $k$ of a binned predictor, and $G,B$ are the corresponding totals. WoE-encoded features and L2 regularization on $\beta$ together preserve the auditable per-feature decomposition that regulators require of a scorecard \citep{Thomas2017}.

\subsection{Gradient boosting on residuals}
\label{ssec:residual}

For observation $i$, the response-based residual is
\begin{equation}
r_i = y_i - \plr(x_i).
\label{eq:residual}
\end{equation}
Pearson or deviance residuals are more standard for GLM diagnostics, but the raw response residual is adopted here because it provides a direct regression target for the secondary learner and aligns with the residual-learning paradigm popularized for deep networks by \citet{He2016}. A gradient boosting model is trained to estimate
\begin{equation}
\hat{r}(x) = f_{\mathrm{GBM}}(x) = \sum_{m=1}^{M} \nu\, h_m(x),
\label{eq:gbm}
\end{equation}
with $h_m$ the regression tree fitted at iteration $m$ and $\nu$ the learning rate. We use XGBoost \citep{ChenGuestrin2016} as the implementation, with LightGBM \citep{Ke2017} retained as a sensitivity check.

\subsection{Hybrid prediction and the interpretability ratio}
\label{ssec:hybrid}

The hybrid prediction additively combines the logistic baseline with the residual learner, weighted by an adaptive coefficient $\alpha^*(x)$ specified in Section~\ref{ssec:adaptive}:
\begin{equation}
\phyb(x) = \mathrm{clip}\!\left(\plr(x) + \alpha^{*}(x)\,\hat{r}(x),\,0,\,1\right).
\label{eq:hybrid}
\end{equation}
To audit the per-prediction contribution of the interpretable branch, we define the \emph{interpretability ratio}
\begin{equation}
\rhox = \frac{|\plr(x)|}{|\plr(x)| + |\alpha^{*}(x)\,\hat{r}(x)|},
\label{eq:rho}
\end{equation}
which is bounded in $[0,1]$ and equals one when the residual correction is zero. Predictions are partitioned into three interpretability regions:
\begin{equation}
\mathrm{Region}(x) = \begin{cases}
\text{Fully interpretable} & \rhox \geq 0.9, \\
\text{Partially interpretable} & 0.7 \leq \rhox < 0.9, \\
\text{ML-driven} & \rhox < 0.7.
\end{cases}
\label{eq:regions}
\end{equation}
The thresholds $0.9$ and $0.7$ are indicative rather than theoretical; their sensitivity is evaluated empirically. Two properties of Eq.~\eqref{eq:rho} should be flagged. First, the ratio is defined on the probability scale rather than the logit scale, so borrowers for whom $\plr(x)$ is large (close to one) tend to fall in the fully-interpretable region when the residual correction $\alpha^{*}(x)\,\hat{r}(x)$ is bounded, which mechanically couples the fully-interpretable region to the high-$\plr$ tail of the score distribution. The empirical finding in Section~\ref{sec:results} that the fully-interpretable region concentrates high-base-rate borrowers is therefore a joint property of Eq.~\eqref{eq:rho} and the well-calibrated behaviour of the logistic-regression baseline on that tail, not a discovered feature independent of the ratio's algebraic form. A logit-scale variant is a natural alternative and is left as future work. Second, the ratio is not invariant under label parity ($y \to 1-y$), since $\plr$ flips accordingly; the interpretation of $\rhox$ is tied to the paper's positive-class convention (default in Taiwan, account ownership in Zindi).

\subsection{Adaptive ensemble weighting}
\label{ssec:adaptive}

The weight $\alpha^{*}(x)$ in Eq.~\eqref{eq:hybrid} is allowed to vary with the local uncertainty of the logistic baseline. Defining the Bernoulli-variance-based uncertainty proxy
\begin{equation}
u_{\mathrm{LR}}(x) = 4 \cdot \plr(x) \cdot (1 - \plr(x)) \in [0, 1],
\label{eq:uncertainty}
\end{equation}
the adaptive weight is
\begin{equation}
\alpha^{*}(x) = \alpha_{\min} + (\alpha_{\max} - \alpha_{\min}) \cdot u_{\mathrm{LR}}(x)^{\gamma},
\label{eq:alpha}
\end{equation}
with $\gamma > 0$ controlling the concavity of the uncertainty-to-weight mapping. The three hyperparameters are selected by grid search over $\alpha_{\min} \in \{0.0, 0.1, 0.2\}$, $\alpha_{\max} \in \{0.3, 0.5, 0.8\}$, $\gamma \in \{0.5, 1.0, 2.0\}$, using 5-fold stratified cross-validation on AUC.

\subsection{Probability calibration}
\label{ssec:cal}

Eq.~\eqref{eq:hybrid} produces probabilities within $[0,1]$ but does not guarantee calibration. We apply Platt scaling post-hoc,
\begin{equation}
p_{\mathrm{cal}}(x) = \frac{1}{1 + \exp\!\left(a \cdot \phyb(x) + b\right)},
\label{eq:platt}
\end{equation}
with $a,b$ estimated on a held-out validation fold. Calibration is reported via reliability diagrams and the Expected Calibration Error
\begin{equation}
\mathrm{ECE} = \sum_{k=1}^{K} \frac{|B_k|}{n}\,\big|\mathrm{acc}(B_k) - \mathrm{conf}(B_k)\big|,
\label{eq:ece}
\end{equation}
where $B_k$ is the $k$-th equal-width probability bin.

\subsection{Prediction-level explanations}
\label{ssec:shap}

For observations in the ML-driven region ($\rhox < 0.7$), the residual term $\alpha^{*}(x)\,\hat{r}(x)$ dominates the final decision and the logistic coefficients no longer suffice as an explanation. We compute TreeSHAP \citep{LundbergLee2017} attributions on $f_{\mathrm{GBM}}$ for these observations, restricted to the residual component to preserve the additive decomposition of Eq.~\eqref{eq:hybrid}. Stability of the SHAP feature rankings is assessed by bootstrap resampling of the test set ($B = 500$), and observed instability is treated as a flag on individual explanations rather than a property of the framework as a whole \citep{AlvarezMelisJaakkola2018, Slack2020}. The Zindi instantiation of this procedure is reported in Section~\ref{ssec:res-shap}.

\subsection{Hard/soft feature-space split}
\label{ssec:hardsoft}

The first methodological extension partitions the Zindi feature set into a \emph{hard} demographic block $X_{\mathrm{hard}}$ (age, gender, marital status, education level, household size, location type, country) and a \emph{soft} behavioural block $X_{\mathrm{soft}}$ (cellphone access, job type, relationship with head of household). The logistic baseline is fit on $X_{\mathrm{hard}}$ only; the residual learner sees $X = X_{\mathrm{hard}} \cup X_{\mathrm{soft}}$. Under this configuration, $\rhox \approx 1$ identifies decisions explainable from hard demographics alone, and $\rhox < 0.7$ identifies decisions in which the soft behavioural proxies materially shifted the outcome. We report this as a methodological note rather than a headline contribution: the soft block on Zindi is modest in width, and a fuller hard/soft separation would require richer alternative-data inputs (mobile-money transaction trails, app behaviour) that this dataset does not include.

\subsection{Fairness audit per interpretability region}
\label{ssec:fairness}

The second extension, which is the empirical focus of the paper, audits the framework's behaviour across protected subgroups at the granularity of the three interpretability regions. For each protected attribute $A$ taking values in $G_A$ (gender $\in \{\mathrm{M}, \mathrm{F}\}$; location $\in \{\mathrm{urban}, \mathrm{rural}\}$; education $\in \{\mathrm{primary~or~less}, \mathrm{secondary}{+}\}$; country $\in \{\mathrm{KE}, \mathrm{RW}, \mathrm{TZ}, \mathrm{UG}\}$), we compute:
\begin{enumerate}
\item \textbf{Region routing rate.} $\Pr[\rhox < 0.7 \mid A = g]$ for each $g \in G_A$, the per-group probability of being routed to the opaque ML-driven region.
\item \textbf{Disparate-impact ratio.} $\mathrm{DI}(g) = \Pr[\hat{y} = 1 \mid A = g] \,/\, \Pr[\hat{y} = 1 \mid A = g_{\mathrm{ref}}]$, with $g_{\mathrm{ref}}$ a designated reference group, evaluated both pooled and restricted to each of the three regions.
\item \textbf{Equalized-odds difference.} $\max_{g}\!\big\{|\mathrm{TPR}(g) - \mathrm{TPR}(g_{\mathrm{ref}})|, |\mathrm{FPR}(g) - \mathrm{FPR}(g_{\mathrm{ref}})|\big\}$ \citep{Hardt2016}, again pooled and per-region.
\end{enumerate}
Each statistic is reported with a 95\% bootstrap confidence interval ($B = 500$ resamples drawn at the observation level). The four-fifths convention $\mathrm{DI} \in [0.8, 1.25]$ (originating in the U.S. EEOC 1978 Uniform Guidelines on Employee Selection Procedures) is reported as a reference threshold rather than a decision rule, since published critiques of the rule are serious and the raw effect sizes are the more informative quantity. The hypothesis under test is not aggregate disparate impact (the existing hybrid literature would not predict that) but subgroup routing to the ML-driven region where adverse-action explanation is weakest, even when aggregate metrics appear balanced.

The per-region variants of the disparate-impact ratio and the equalized-odds difference in list items~2--3 above condition on $\rhox$, which is causally downstream of $X$ and therefore a post-treatment variable in the sense of \citet{Rosenbaum1984}. Region-conditional DI and EOD are consequently descriptive stratifications of the audit surface rather than fairness metrics defined on well-behaved subpopulations, and their per-region values should not be interpreted as ``fairness within the region.'' The routing-rate audit (list item~1) is not affected, because $\rhox$ is the outcome rather than a conditioning variable.

\subsection{Thin-file segmentation}
\label{ssec:thinfile}

The third extension defines an operational thin-file subgroup on the Zindi feature set as the bottom quartile of a composite index combining education level, cellphone access, and formal-job status. On this subgroup we report (a) standalone LR, standalone XGBoost, and hybrid AUC with bootstrap CIs on the pairwise $\Delta\mathrm{AUC}$; (b) the distribution of $\rhox$ on the subgroup, to test whether the framework relies more heavily on the residual learner for thin-file borrowers; and (c) calibration metrics (Brier, ECE) restricted to the subgroup.

\subsection{Joint optimization via alternating minimization}
\label{ssec:jointopt}

The training procedure in Sections~\ref{ssec:lr}--\ref{ssec:residual} is two-stage: $\beta$ is estimated first, then $f_{\mathrm{GBM}}$ is fitted on the residuals produced under that fixed $\beta$. The fourth extension formulates the joint loss
\begin{equation}
\mathcal{L}(\beta, f_{\mathrm{GBM}}) = \sum_{i=1}^{N} \ell\!\left(y_i, \mathrm{clip}(\plr(x_i; \beta) + \alpha^{*}(x_i)\,f_{\mathrm{GBM}}(x_i), 0, 1)\right) + \lambda_{\mathrm{LR}}\,\|\beta\|_2^{2} + \Omega(f_{\mathrm{GBM}}),
\label{eq:jointloss}
\end{equation}
where $\ell$ is the binary cross-entropy and $\Omega$ is the XGBoost complexity penalty. The procedure implemented here is an alternating fixed-point iteration on the two-stage recipe rather than exact joint minimization of Eq.~\eqref{eq:jointloss}: (i) hold $f_{\mathrm{GBM}}$ fixed and update $\beta$ by L-BFGS on $\mathcal{L}$; (ii) hold $\beta$ fixed and refit $f_{\mathrm{GBM}}$ on the response residuals under the updated $\beta$; (iii) iterate until the change in held-out AUC falls below a tolerance. Step (ii) minimizes squared error on the residuals rather than the joint binary cross-entropy in Eq.~\eqref{eq:jointloss}, so this scheme is not coordinate descent on $\mathcal{L}$; a true functional-gradient descent on the joint loss (which would flow gradients through both branches) is left as future work. We report convergence trajectory and the final $\Delta\mathrm{AUC}$ against the two-stage training of the thesis. Implementation is described in Section~\ref{sec:exp} and is publicly available alongside the paper.

\section{Datasets and the Zindi reframing}
\label{sec:datasets}

We evaluate the framework on two primary datasets and one conditional robustness dataset. The Zindi Financial Inclusion in Africa dataset \citep{ZindiFSD2018} carries the African empirical contribution and supports the fairness and thin-file analyses of Sections~\ref{ssec:fairness}--\ref{ssec:thinfile}. The Taiwan Credit Default dataset \citep{YehLien2009}, used in the methodological work of \citet{Kanziga2026thesis}, is retained as a benchmark for continuity. The Kiva loans dataset, restricted to its African subset, is used conditionally for a robustness check on actual loan-adverse outcomes.

\subsection{Zindi Financial Services Deepening dataset}
\label{ssec:zindi}

The data are drawn from FinScope household surveys conducted by the Financial Sector Deepening network across Kenya, Rwanda, Tanzania, and Uganda, and were released as the training set of the Zindi Financial Inclusion in Africa challenge \citep{ZindiFSD2018}. The labelled training set contains 23{,}524 respondents spanning survey years 2016--2018 (Rwanda 8{,}735; Tanzania 6{,}620; Kenya 6{,}068; Uganda 2{,}101), with a marginal positive-class rate of 14.08\% (account ownership). The variables comprise basic demographic descriptors (age, gender, marital status, education level, household size), geographic and infrastructure indicators (country, location type, cellphone access), and labour-market status (job type, relationship with head of household). The label is \texttt{bank\_account}, a binary indicator of whether the respondent owns an account at a formal financial institution. The dataset has zero missing values, simplifying preprocessing relative to Taiwan.

\subsection{Target reframing}
\label{ssec:reframing}

The Zindi label is not a default outcome. We use it as a deliberate proxy for the upstream eligibility decision that an African digital lender faces when selecting which thin-file applicants to underwrite at all. Three considerations support this reframing.

First, in the formal-bank-account-poor populations of East Africa \citep{Findex2021}, the binary distinction between borrowers a regulated lender can score from bureau data and those it cannot is precisely the distinction the \texttt{bank\_account} label captures. The accuracy-interpretability-fairness trade-off facing a lender on this decision is structurally identical to the trade-off on default prediction: a model classifies applicants, a regulator audits, and protected subgroups may be disadvantaged by the model's opacity.

Second, the protected-attribute coverage of the Zindi data (gender, urban/rural, education, country) is, to our reading, unmatched by any publicly available African default-labelled dataset. Restricting the analysis to default outcomes would force the African empirical evidence onto datasets without the fairness coverage that the paper's central question requires.

Third, the methodological apparatus of the framework (LR baseline, residual learner, $\rhox$ decomposition, fairness audit per interpretability region) does not depend on the semantics of the label. It depends on the existence of a binary classification problem with protected attributes, and on the lender's interest in auditable per-prediction explanations. Both conditions hold under the reframing.

We acknowledge two limitations directly. First, the magnitude of the fairness violations measured on Zindi cannot be transferred uncritically to a default-prediction setting on the same population, because the conditional distribution of the label given protected attributes differs between account ownership and default. Second, and more consequentially, the two labels sit on opposite sides of the lender's selection funnel: bank-account ownership is a cross-sectional outcome jointly shaped by supply-side factors (branch coverage, agent networks, KYC infrastructure) and demand-side factors, whereas default is measured only on borrowers a lender has already selected in. A rural respondent without a bank account may have been screened out at the eligibility gate by supply factors that have no analog in default prediction. The Zindi analysis should therefore be read as an audit of the \emph{eligibility gate} rather than as a proxy for approved-borrower default risk; the two are complementary but not interchangeable. The contribution of the empirical analysis is the demonstration that the proposed audit (a) is operationally feasible, (b) detects subgroup routing into the opaque region when it occurs, and (c) yields actionable findings at the per-prediction level for the population-selection decision. The Kiva subset (Section~\ref{ssec:kiva}) was originally scoped as a robustness check on actual loan-adverse-outcome labels for the approved-borrower side of the funnel, on a smaller population for which protected-attribute coverage is weaker; that check is left to a follow-up paper (Section~\ref{ssec:disc-westafrica}).

\subsection{Taiwan Credit Default dataset}
\label{ssec:taiwan}

The Taiwan Credit Default dataset \citep{YehLien2009}, available from the UCI repository \citep{DuaGraff2019}, contains 30{,}000 credit card holders with 23 features (demographics, six months of payment history, six months of bill amounts, six months of payment amounts) and a binary default-in-the-next-month label. The class imbalance is approximately 22\% positive. The dataset is retained from the thesis without alteration of preprocessing, to permit direct numerical continuity with the headline results of \citet{Kanziga2026thesis} (Section~\ref{sec:results}). Protected attributes are not analysed on this dataset: the available demographic columns (sex, education, marital status) are coarse and Taiwan-specific, and the regulatory framing of the paper does not extend to that jurisdiction.

\subsection{Kiva African subset (conditional robustness)}
\label{ssec:kiva}

The Kiva platform's public loan-level data include disbursement, repayment, and loss outcomes on microloans distributed through field partners across multiple African countries, and were originally scoped as a robustness check on actual loan-adverse-outcome labels rather than the account-ownership proxy of Zindi. The present paper does not include the Kiva analysis: under the project's risk register, the Kiva robustness was the first item to be cut when the timeline compressed, and the empirical contribution rests on Zindi alone. A follow-up paper or extension report would naturally include the Kiva analysis as the first piece of additional evidence; the audit pipeline of Section~\ref{ssec:fairnesspipe} is dataset-agnostic and applies to Kiva without modification beyond the field-partner-specific protected-attribute definitions.

\subsection{Operational definitions: protected attributes and thin-file status}
\label{ssec:opdefs}

For the fairness analysis on Zindi, protected attributes are coded as follows. \emph{Gender}: as recorded. \emph{Location}: urban vs.\ rural per the FinScope survey instrument. \emph{Education}: primary or less vs.\ secondary or above, collapsing the survey's finer categories to ensure adequate per-cell counts. \emph{Country}: Kenya, Rwanda, Tanzania, Uganda, with Kenya as the reference group in the main analysis (the largest single-country subsample; per-country sensitivity to reference choice is left to future work). The thin-file subgroup of Section~\ref{ssec:thinfile} is defined as the bottom quartile of a composite index obtained by summing the standardized values of education level, cellphone access (binary), and a formal-job indicator (binary). The quartile threshold is computed on the full Zindi training set and applied identically to validation and test partitions.

\section{Experimental setup}
\label{sec:exp}

This section specifies preprocessing, model configurations, the training and evaluation pipeline, the fairness audit module, and the bootstrap inference protocol. The pipeline that produced the Taiwan results retained from \citet{Kanziga2026thesis} is preserved without modification to permit direct numerical comparison; the Zindi pipeline mirrors it except where dataset-specific decisions are noted.

\subsection{Preprocessing}
\label{ssec:preproc}

Missing values are imputed by median (numerical) and mode (categorical) on the training partition only, and the imputation values are then applied to validation and test. Rare WoE values encountered in held-out folds that do not appear in training are assigned a neutral $\mathrm{WoE} = 0$, preserving the log-odds-zero contribution and preventing information leakage. Numerical features are winsorized at the 1st and 99th percentiles of the training partition. For the logistic regression component, categorical features are transformed via WoE encoding after monotonic binning of continuous predictors; for the residual learner, categorical features are one-hot encoded, which permits the boosting model to capture nonlinear interactions among categorical levels. Feature selection for the logistic regression component uses an Information Value threshold of $\mathrm{IV} \geq 0.02$. Class imbalance is addressed via balanced class weighting at both the logistic regression and the XGBoost level; an additional SMOTE comparison on the WoE-transformed feature space, reported in \citet{Kanziga2026thesis} (Section~5.6), showed that synthetic oversampling provides no meaningful AUC improvement over class weighting on these datasets, and class weighting is therefore retained throughout.

\begin{sloppypar}
For Zindi specifically, all ten input features (\texttt{country}, \texttt{location\_type}, \texttt{cellphone\_access}, \texttt{gender\_of\_respondent}, \texttt{relationship\_with\_head}, \texttt{marital\_status}, \texttt{education\_level}, \texttt{job\_type}, \texttt{household\_size}, \texttt{age\_of\_respondent}) survive the $\mathrm{IV} \geq 0.02$ filter; the strongest predictor by IV is \texttt{education\_level} ($\mathrm{IV} = 0.97$), followed by \texttt{cellphone\_access} ($0.71$) and \texttt{job\_type} ($0.63$). The \texttt{year} column (2016--2018) and the \texttt{uniqueid} identifier are dropped prior to fitting: \texttt{year} because it is perfectly collinear with \texttt{country} in this sample (each survey year covers one country), and \texttt{uniqueid} because it is a unique-per-row identifier with no predictive content. The training partition contains $14{,}114$ observations after a 60/20/20 stratified split.
\end{sloppypar}

\subsection{Model configurations}
\label{ssec:configs}

\begin{sloppypar}
Table~\ref{tab:configs} summarizes the model configurations. All settings are inherited from \citet{Kanziga2026thesis} (Table~4.2) on the Taiwan dataset and applied unaltered on Zindi, with the exception of \texttt{scale\_pos\_weight} which is recomputed from the Zindi class balance.
\end{sloppypar}

\begin{table}[t]
\centering
\caption{Final model configurations. Inherited from \citet{Kanziga2026thesis}, Table~4.2.}
\label{tab:configs}
\footnotesize
\begin{tabular}{lll}
\toprule
Model & Parameter & Value \\
\midrule
\multirow{4}{*}{Logistic regression}
& Solver & \texttt{lbfgs} \\
& Regularization & L2 \\
& Class weight & balanced \\
& $C$ grid & $\{0.01, 0.1, 1, 10\}$ \\
\midrule
\multirow{8}{*}{Residual XGBoost}
& Objective & \texttt{reg:squarederror} \\
& Learning target & raw residuals $r_i = y_i - \plr(x_i)$ \\
& Estimators & 300--500 \\
& Max depth & 3--4 \\
& Learning rate & 0.03--0.05 \\
& Subsample & 0.8 \\
& \texttt{colsample\_bytree} & 0.8 \\
& Regularization $\lambda$ & 2 \\
\midrule
\multirow{5}{*}{Standalone XGBoost}
& Estimators & 400 \\
& Max depth & 4 \\
& Learning rate & 0.03 \\
& Subsample & 0.8 \\
& \texttt{colsample\_bytree} & 0.8 \\
\midrule
\multirow{6}{*}{Hybrid (adaptive)}
& Residual definition & raw residual \\
& Static $\alpha$ grid & $\{0.0, 0.1, \ldots, 1.0\}$ \\
& $\alpha_{\min}$ grid & $\{0.00, 0.10, 0.20\}$ \\
& $\alpha_{\max}$ grid & $\{0.30, 0.50, 0.80\}$ \\
& $\gamma$ grid & $\{0.5, 1.0, 2.0\}$ \\
& Calibration & Platt scaling on validation logits \\
\midrule
Bootstrap & Resamples & 500 (stratified) \\
\bottomrule
\end{tabular}
\end{table}

The shallow tree depths (3--4) for the residual learner are deliberate: deeper trees allow the boosting component to dominate the prediction and erode the interpretability ratio. Low learning rates (0.03--0.05) encourage gradual residual correction rather than aggressive nonlinear replacement of the logistic baseline. Static and adaptive weighting hyperparameters are selected via 5-fold and 3-fold stratified cross-validation respectively, with validation AUC as the optimization criterion.

\subsection{Training and evaluation protocol}
\label{ssec:protocol}

To prevent information leakage, all preprocessing, hyperparameter tuning, residual-learning, and Platt-calibration stages are performed exclusively on the training and validation partitions. The test partition is reserved strictly for final evaluation. Random seeds are fixed across runs for reproducibility. Implementation uses scikit-learn, XGBoost, and standard Python scientific libraries; the fairness audit module is implemented as part of the open-source release described in Section~\ref{sec:codedata}.

\subsection{Evaluation metrics}
\label{ssec:metrics}

Discrimination is evaluated via the Area Under the ROC Curve (AUC), the Gini coefficient ($2\,\mathrm{AUC} - 1$), and the minority-class F1-score. Calibration is evaluated via the Brier score and the Expected Calibration Error of Eq.~\eqref{eq:ece} with $K = 10$ equal-width probability bins, complemented by reliability diagrams. Interpretability behaviour is summarized via the distribution of $\rhox$ on the test partition and the share of observations in each of the three interpretability regions of Eq.~\eqref{eq:regions}, conditional on the empirical default rate (Taiwan) or adverse-eligibility rate (Zindi) within each region.

\subsection{Fairness audit pipeline}
\label{ssec:fairnesspipe}

For each protected attribute $A$ defined in Section~\ref{ssec:opdefs}, the audit module computes the three statistics introduced in Section~\ref{ssec:fairness}: per-group region routing rate, disparate-impact ratio, and equalized-odds difference. Each statistic is computed pooled (over all test observations) and stratified by interpretability region. The stratified estimates are the central quantities of interest: the hypothesis under test is not that the framework produces aggregate disparate impact but that protected subgroups may be disproportionately routed into the ML-driven region while aggregate metrics appear balanced.

The reference group $g_{\mathrm{ref}}$ is set to the majority subgroup on each attribute (male for gender, urban for location, secondary-and-above for education, Kenya for country), with sensitivity to that choice reported as a robustness check.

\subsection{Bootstrap statistical inference}
\label{ssec:bootstrap}

Confidence intervals on all AUC differences, disparate-impact ratios, and equalized-odds differences are constructed by stratified paired bootstrap on the held-out test partition. Stratification samples with replacement within the positive-class and negative-class strata separately (preserving the marginal class distribution within each resample); for quantities conditioned on a protected attribute we sample within (class $\times$ attribute) strata. For AUC-difference intervals we use $B = 2000$ resamples; for the fairness quantities (routing rate, disparate impact, equalized-odds difference) we use $B = 500$. Statistical significance of pairwise AUC differences (Hybrid vs.\ LR, Hybrid vs.\ XGBoost) is assessed by the paired DeLong test on correlated ROC curves, using the fast midrank implementation of \citet{SunXu2014}; the DeLong test provides an asymptotic $z$-statistic and $p$-value that are not floor-bounded by the bootstrap resample count. Both a one-sided $p$-value (H$_1$: hybrid AUC $>$ comparator AUC) and a two-sided $p$-value are reported. The bootstrap CI is retained as the primary interval statement; the DeLong test is the primary hypothesis test. Both are reported in Table~\ref{tab:bootstrap}. Consistent with the descriptive-audit intent of Section~\ref{ssec:fairness}, we do not adjust for multiple comparisons across the four protected attributes and three regions, and we do not treat the region-stratified fairness quantities as confirmatory tests.

\section{Results}
\label{sec:results}

This section reports results addressing the four research questions of Section~\ref{sec:intro}. Taiwan results are inherited from \citet{Kanziga2026thesis} and serve as the methodological benchmark; the Zindi-specific results are the empirical contribution of the present paper.

\subsection{Predictive discrimination (RQ1)}
\label{ssec:res-rq1}

Table~\ref{tab:perf} reports test-set predictive performance.

\begin{table}[t]
\centering
\caption{Test-set predictive performance. Taiwan results are reproduced from \citet{Kanziga2026thesis} (Table~5.1); Zindi results are produced by the present paper.}
\label{tab:perf}
\small
\begin{tabular}{llcccc}
\toprule
Dataset & Model & AUC & Gini & Brier & F1 \\
\midrule
\multirow{4}{*}{Taiwan Credit}
& Logistic regression & 0.719 & 0.438 & 0.207 & 0.468 \\
& XGBoost & 0.775 & 0.550 & 0.177 & \textbf{0.538} \\
& Hybrid (uncalibrated) & 0.776 & 0.551 & 0.139 & 0.495 \\
& Hybrid (Platt calibrated) & \textbf{0.776} & \textbf{0.552} & \textbf{0.136} & 0.470 \\
\midrule
\multirow{5}{*}{Zindi FSD}
& Logistic regression & 0.854 & 0.708 & 0.158 & 0.493 \\
& LR + Platt (ablation) & 0.854 & 0.708 & 0.088 & 0.466 \\
& XGBoost & \textbf{0.873} & \textbf{0.746} & 0.143 & 0.513 \\
& Hybrid (uncalibrated) & 0.869 & 0.738 & 0.092 & \textbf{0.529} \\
& Hybrid (Platt calibrated) & 0.869 & 0.738 & \textbf{0.085} & 0.499 \\
\bottomrule
\end{tabular}
\end{table}

On the Taiwan dataset, the hybrid model after Platt calibration achieves AUC $= 0.776$, matching the standalone XGBoost benchmark ($\Delta\mathrm{AUC} = +0.001$, 95\% CI $[-0.004, +0.006]$, $p = 0.37$) while improving over standalone logistic regression by 5.7 AUC points ($\Delta\mathrm{AUC} = +0.057$, 95\% CI $[+0.048, +0.066]$, $p < 0.001$). The Brier score is reduced from $0.177$ for standalone XGBoost to $0.136$ for the calibrated hybrid, a relative reduction of approximately 23\%, accompanied by a reduction in Expected Calibration Error from $0.234$ (LR) to $0.011$ (calibrated hybrid).

On the Zindi data, the framework's behaviour replicates the Taiwan qualitative pattern. The calibrated hybrid achieves AUC $= 0.869$ on the held-out test partition ($n_{\mathrm{test}} = 4{,}705$, positive rate 14.07\%), matching standalone XGBoost ($\Delta\mathrm{AUC} = -0.004$, 95\% CI $[-0.009, +0.001]$, $p = 0.92$) and improving on standalone logistic regression by 1.5 AUC points ($\Delta\mathrm{AUC} = +0.015$, 95\% CI $[+0.011, +0.020]$, $p < 0.001$). The Brier score drops from $0.158$ (LR) to $0.085$ (calibrated hybrid), a relative reduction of approximately 46\%, larger than the Taiwan reduction of 23\%; the LR baseline is more severely miscalibrated on the highly imbalanced Zindi target (ECE $= 0.234$ for LR; $0.024$ for the calibrated hybrid, a tenfold reduction). The LR + Platt ablation row in Table~\ref{tab:perf} attributes the calibration gain: Platt scaling on the LR baseline alone recovers Brier $0.088$ and ECE $0.016$, i.e., most of the Brier reduction and essentially all of the ECE reduction observed for the calibrated hybrid come from Platt calibration of the class-weighted baseline rather than from the residual reshaping. The residual learner contributes the discrimination gain ($+0.015$ AUC vs.\ LR), not the calibration gain. The adaptive-weighting hyperparameters selected by cross-validation on Zindi are $\alpha_{\min} = 0.2$, $\alpha_{\max} = 0.8$, $\gamma = 1.0$, qualitatively close to the Taiwan choices.

The F1-score asymmetry visible in Table~\ref{tab:perf} for the Taiwan dataset (calibrated hybrid lower than standalone XGBoost despite equal AUC) is a consequence of the Platt-scaling shift of predicted probabilities toward the empirical default rate, which reduces the share of observations exceeding the $0.5$ classification threshold. In practical deployment the classification threshold is typically recalibrated using expected-cost optimization or Youden's $J$ statistic rather than relying on $0.5$; F1 at $0.5$ is reported for comparison with the methodological literature, not as the operational decision rule.

\begin{figure}[t]
\centering
\includegraphics[width=0.62\linewidth]{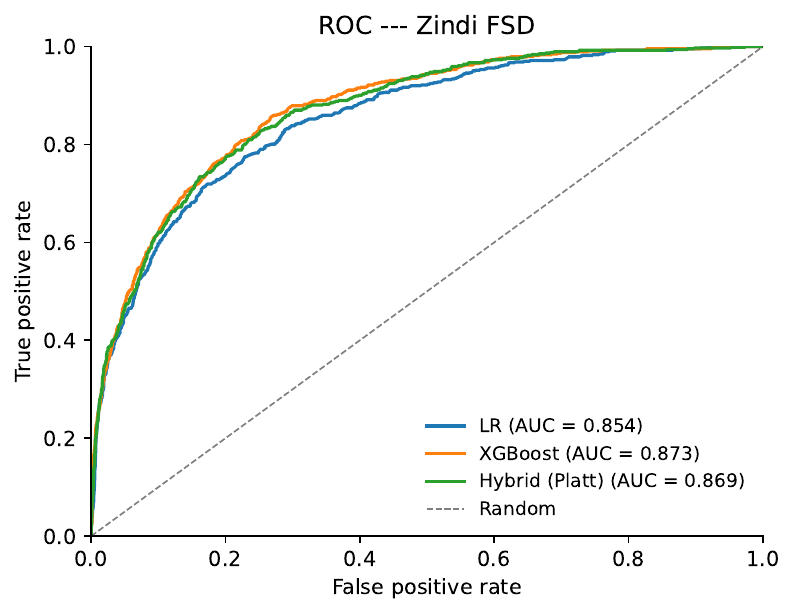}
\hfill
\includegraphics[width=0.62\linewidth]{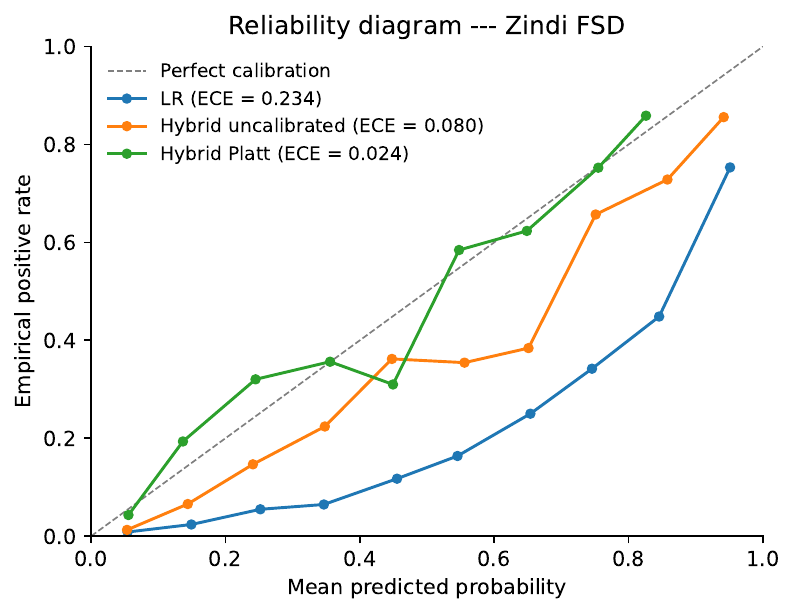}
\caption{Zindi FSD test-set diagnostics. \emph{Left:} ROC curves for the three competing models. The calibrated hybrid tracks the standalone XGBoost curve over essentially the entire range and dominates the LR baseline. \emph{Right:} reliability diagram with Expected Calibration Error. Platt calibration brings the hybrid close to the diagonal; the standalone LR is severely miscalibrated (ECE $= 0.234$) under balanced class weighting.}
\label{fig:zindi-discrimination}
\end{figure}

Figure~\ref{fig:zindi-discrimination} displays the ROC curves and reliability diagrams underlying these statistics.

\subsection{Interpretability region behaviour (RQ2)}
\label{ssec:res-rq2}

Table~\ref{tab:regions} reports the distribution of test-set observations across the three interpretability regions of Eq.~\eqref{eq:regions}, together with the empirical default (or adverse-eligibility) rate and the LR and Hybrid AUC computed within each region.

\begin{table}[t]
\centering
\caption{Distribution of test-set observations across interpretability regions, with within-region default rate and per-region AUC. Taiwan results from \citet{Kanziga2026thesis} (Table~5.5); Zindi results from this paper.}
\label{tab:regions}
\resizebox{\textwidth}{!}{%
\begin{tabular}{llcccc}
\toprule
Dataset & Region & Share (\%) & Default rate (\%) & AUC (LR) & AUC (Hybrid) \\
\midrule
\multirow{3}{*}{Taiwan Credit}
& Fully interpretable ($\rho \geq 0.9$) & 6.3 & \textbf{69.5} & 0.653 & 0.669 \\
& Partially interpretable & 28.6 & 33.4 & 0.643 & \textbf{0.698} \\
& ML-driven ($\rho < 0.7$) & \textbf{65.1} & 12.5 & 0.601 & 0.670 \\
\midrule
\multirow{3}{*}{Zindi FSD}
& Fully interpretable ($\rho \geq 0.9$) & 6.4 & \textbf{53.8} & 0.869 & 0.888 \\
& Partially interpretable & 34.1 & 19.7 & 0.821 & 0.830 \\
& ML-driven ($\rho < 0.7$) & \textbf{59.4} & 6.5 & 0.730 & 0.752 \\
\bottomrule
\end{tabular}%
}
\end{table}

The Taiwan results exhibit the central qualitative pattern of the framework: the fully interpretable region, though small in share (6.3\%), concentrates the highest-default-rate borrowers (69.5\%). As flagged in Section~\ref{ssec:hybrid}, this concentration is a joint consequence of the probability-scale definition of $\rho$ in Eq.~\eqref{eq:rho} (which places high-$\plr$ borrowers in the fully-interpretable region) and the well-calibrated behaviour of the logistic-regression baseline on the high-risk tail, rather than an independent empirical property of the framework. The operationally relevant reading is therefore that high-stakes lending decisions remain primarily governed by the auditable logistic-regression scorecard on both datasets, while the ML-driven region (larger in share at 65.1\%, but lower in default rate at 12.5\%) accommodates the bulk of the population in which the residual learner contributes most. Per-region AUC is itself informative: the hybrid model uplifts AUC in the partially interpretable region (LR $0.643 \to$ Hybrid $0.698$) and in the ML-driven region (LR $0.601 \to$ Hybrid $0.670$) where the residual correction has the most room to operate.

On the Zindi data, the same qualitative pattern obtains: the fully interpretable region, though small in share (6.4\%), concentrates respondents with the highest positive-class rate (53.8\% account ownership, against a population marginal of 14.1\%), while the ML-driven region contains the majority of test observations (59.4\%) at a much lower positive rate (6.5\%). The per-region AUC gain from LR to Hybrid is largest in the ML-driven region ($0.730 \to 0.752$, $+0.022$), consistent with the framework's design: the residual learner contributes most where the linear baseline is least certain. The structural alignment between the two datasets is the central finding for RQ2: the framework concentrates high-stakes (high-base-rate) decisions in the auditable region across two datasets with very different label semantics (default prediction in Taiwan, account ownership in East Africa), which suggests the routing property is a stable feature of the framework rather than a Taiwan-specific artifact.

\begin{figure}[t]
\centering
\includegraphics[width=0.7\linewidth]{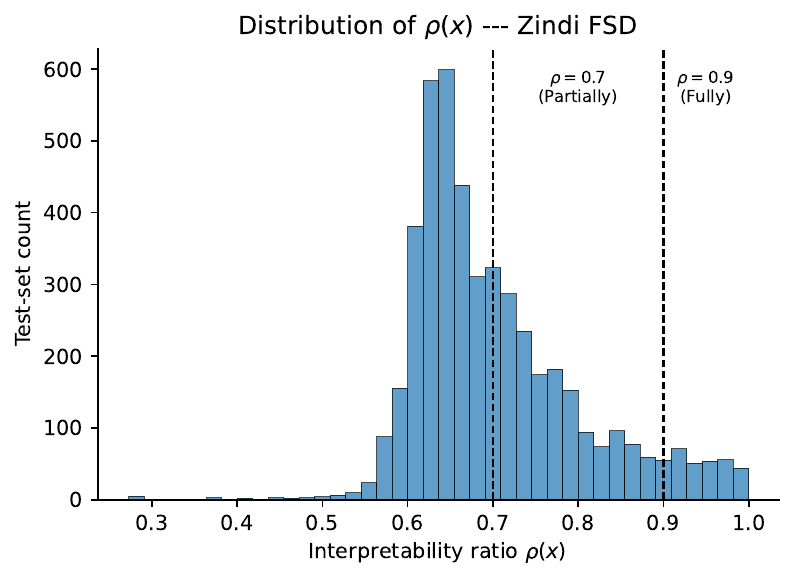}
\caption{Test-set distribution of the interpretability ratio $\rho(x)$ on Zindi FSD. Dashed lines mark the three-region thresholds $\rho = 0.9$ (Fully) and $\rho = 0.7$ (Partially). The distribution is bimodal: a primary mode below the ML-driven threshold and a secondary mode near $\rho = 1$, with the partially-interpretable region between them as a transition band.}
\label{fig:zindi-rho}
\end{figure}

\subsection{Fairness audit per interpretability region (RQ3)}
\label{ssec:res-rq3}

\begin{figure}[t]
\centering
\includegraphics[width=0.78\linewidth]{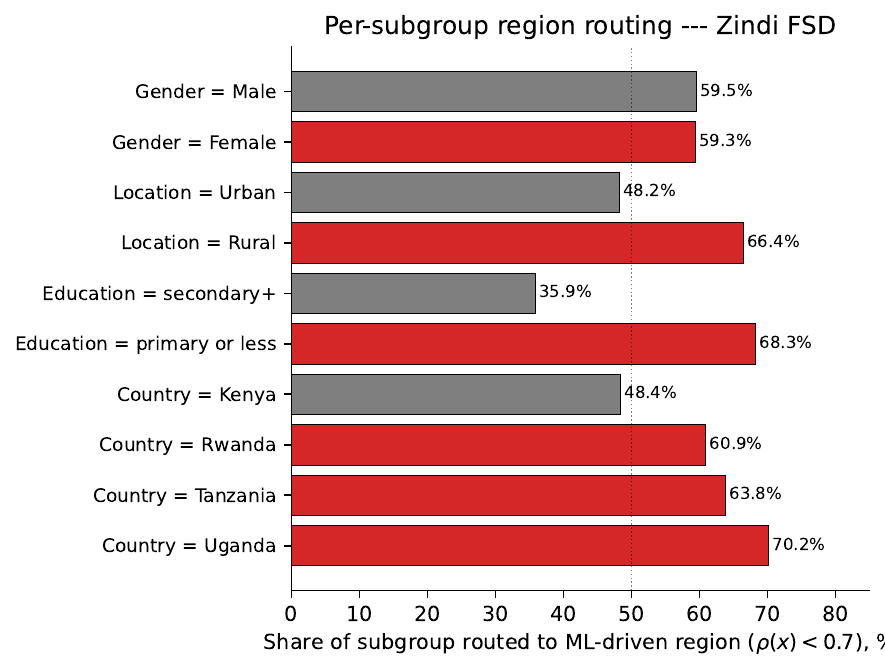}
\caption{Share of each protected subgroup routed to the ML-driven interpretability region ($\rho(x) < 0.7$) on the Zindi test set. Reference groups (Male, Urban, secondary-and-above, Kenya) are shown in grey; comparison groups in red. Gender shows essentially no routing disparity; location, education, and country exhibit gaps of 18, 32, and up to 22 percentage points respectively.}
\label{fig:zindi-fairness}
\end{figure}

The framework's three-region partition of test-set predictions ($6.4\%$ fully interpretable, $34.1\%$ partially, $59.4\%$ ML-driven) sorts borrowers, and the central empirical question of the paper is whether that sorting is independent of protected attributes. It is not. Figure~\ref{fig:zindi-fairness} displays the headline pattern: rural, primary-or-less-educated, and Ugandan respondents are routed to the ML-driven region at rates 18, 32, and 22 percentage points above their respective reference subgroups, while gender shows essentially no routing disparity. Tables~\ref{tab:fairness-routing}, \ref{tab:fairness-di}, and \ref{tab:fairness-eod} report the per-region routing rate, disparate-impact ratio, and equalized-odds difference for each protected attribute on the Zindi test set; the routing rates in Table~\ref{tab:fairness-routing} are the quantity to read first.

\begin{table}[t]
\centering
\caption{Region-routing rate by protected subgroup on the Zindi test set: $\Pr[\rho(x) < 0.7 \mid A = g]$ with 95\% bootstrap confidence intervals ($B = 500$).}
\label{tab:fairness-routing}
\small
\begin{tabular}{lccc}
\toprule
Subgroup & $n$ & $\Pr[\rho < 0.7]$ & 95\% CI \\
\midrule
Gender $=$ male (ref) & 1{,}913 & 0.595 & $[0.577,\,0.618]$ \\
Gender $=$ female & 2{,}792 & 0.594 & $[0.576,\,0.613]$ \\
Location $=$ urban (ref) & 1{,}809 & 0.482 & $[0.458,\,0.505]$ \\
Location $=$ rural & 2{,}896 & \textbf{0.664} & $[0.649,\,0.684]$ \\
Education $=$ secondary$+$ (ref) & 1{,}283 & 0.359 & $[0.332,\,0.385]$ \\
Education $=$ primary or less & 3{,}422 & \textbf{0.683} & $[0.669,\,0.698]$ \\
Country $=$ Kenya (ref) & 1{,}209 & 0.484 & $[0.456,\,0.516]$ \\
Country $=$ Rwanda & 1{,}746 & 0.609 & $[0.588,\,0.632]$ \\
Country $=$ Tanzania & 1{,}277 & 0.638 & $[0.614,\,0.666]$ \\
Country $=$ Uganda & 473 & \textbf{0.702} & $[0.663,\,0.742]$ \\
\bottomrule
\end{tabular}
\end{table}

\begin{table}[t]
\centering
\caption{Disparate-impact ratio $\mathrm{DI}(g) = \Pr[\hat{y} = 1 \mid A = g] \,/\, \Pr[\hat{y} = 1 \mid A = g_{\mathrm{ref}}]$ on the Zindi test set, pooled and stratified by interpretability region (decision threshold $0.5$). Four-fifths reference: $\mathrm{DI} \in [0.8, 1.25]$. ``---'' in the ML-driven column denotes that the reference group has no predicted positives in that region (vanishing denominator).}
\label{tab:fairness-di}
\small
\begin{tabular}{lcccc}
\toprule
Subgroup (vs.\ reference) & Pooled DI & Fully & Partially & ML-driven \\
\midrule
Gender $=$ female (vs.\ male) & 0.429 & 0.530 & 0.409 & --- \\
Location $=$ rural (vs.\ urban) & 0.309 & 0.585 & 0.487 & --- \\
Education $=$ primary or less (vs.\ secondary$+$) & \textbf{0.025} & 0.056 & 0.075 & --- \\
Country $=$ Rwanda (vs.\ Kenya) & 0.190 & 0.544 & 0.172 & --- \\
Country $=$ Tanzania (vs.\ Kenya) & 0.291 & 0.598 & 0.260 & --- \\
Country $=$ Uganda (vs.\ Kenya) & 0.216 & 0.592 & 0.370 & --- \\
\bottomrule
\end{tabular}
\end{table}

\begin{table}[t]
\centering
\caption{Equalized-odds difference $\max_{g}\{|\mathrm{TPR}(g) - \mathrm{TPR}(g_{\mathrm{ref}})|, |\mathrm{FPR}(g) - \mathrm{FPR}(g_{\mathrm{ref}})|\}$ on the Zindi test set, pooled and per region \citep{Hardt2016}. ``---'' in the ML-driven column denotes that the reference group has no predicted positives in that region, so the TPR/FPR denominators collapse and EOD is undefined (not zero); reported explicitly to avoid the false implication of perfect equalized odds.}
\label{tab:fairness-eod}
\small
\begin{tabular}{lcccc}
\toprule
Attribute (reference) & Pooled & Fully & Partially & ML-driven \\
\midrule
Gender (male) & 0.104 & 0.437 & 0.097 & --- \\
Location (urban) & 0.238 & 0.270 & 0.101 & --- \\
Education (secondary$+$) & \textbf{0.537} & 0.860 & 0.319 & --- \\
Country (Kenya) & 0.266 & 0.651 & 0.282 & --- \\
\bottomrule
\end{tabular}
\end{table}

The audit produces four findings.

\textit{Routing disparity along socioeconomic axes is large; along gender it is essentially zero.} Table~\ref{tab:fairness-routing} shows that rural respondents are routed to the ML-driven region at 66.4\% versus 48.2\% for urban (a gap of 18 percentage points, with non-overlapping bootstrap confidence intervals), primary-or-less-education respondents at 68.3\% versus 35.9\% for secondary-and-above (a gap of 32 percentage points), and Ugandan respondents at 70.2\% versus 48.4\% for Kenyan (22 percentage points). The cross-attribute pattern matters: the routing disparity tracks markers of socioeconomic disadvantage (rural, low-education, smaller-market country) rather than tracking gender, which exhibits only a 0.1-percentage-point male--female difference. This pattern is consistent with the framework's design: where the logistic-regression baseline has higher uncertainty (low base rate, sparse features), the residual learner contributes more. The populations on which the baseline is least confident are, in this data, the populations with the thinnest formal-financial-system footprint.

\textit{Aggregate disparate impact fails the four-fifths rule on every protected attribute except gender, but the magnitude is largely driven by baseline inequality.} The pooled DI ratios in Table~\ref{tab:fairness-di} are well outside $[0.8, 1.25]$ for location ($0.309$), country ($0.190$--$0.291$), and most starkly for education ($0.025$). The education number deserves explicit interpretation: primary-or-less-educated respondents are predicted to be in the formal banking system at 2.5\% the rate of secondary-or-above respondents. This is a model-level reflection of the underlying population-level disparity. The model is not creating this inequality; it is encoding the structural inequality already present in formal financial access in the survey populations. A regulator using this audit on a deployed lender would correctly conclude that the audit catches both model-introduced disparity and baseline structural disparity, and would need additional methodology to disentangle them.

\textit{Per-region DI and EOD stratifications are descriptive, not causal.} As noted in Section~\ref{ssec:fairness}, conditioning on $\rhox$ conditions on a post-treatment variable causally downstream of $X$ \citep{Rosenbaum1984}. The per-region rows in Tables~\ref{tab:fairness-di} and~\ref{tab:fairness-eod} should therefore be read as a stratified profile of the audit surface, not as ``fairness within the region.'' Two patterns nonetheless carry descriptive interest. First, the ML-driven region columns in Table~\ref{tab:fairness-di} show DI ratios at or near zero across all attributes because almost no observations in that region are predicted as positive (the residual correction is small and base rates there are low), so the denominator collapses; the entries should be read as undefined rather than as evidence of fairness. Second, in the Fully and Partially regions the pooled EOD ordering education ($0.537$) $>$ country ($0.266$) $>$ location ($0.238$) $>$ gender ($0.104$) obtains in Table~\ref{tab:fairness-eod}, with the Fully region carrying the largest EOD values; the region in which the logistic-regression scorecard exerts the most influence is also the region where the TPR/FPR gap by protected group is widest, which is consistent with the routing finding that the framework offloads decisions on disadvantaged subgroups to the residual learner and keeps decisions on the reference subgroups in the interpretable branch.

\subsection{Thin-file segmentation}
\label{ssec:res-thinfile}

Table~\ref{tab:thinfile} reports predictive performance restricted to the thin-file subgroup of Section~\ref{ssec:thinfile} on the Zindi data.

\begin{table}[t]
\centering
\caption{Performance on the thin-file subgroup (bottom-quartile composite index of education, cellphone access, and formal-job indicator) on the Zindi test set.}
\label{tab:thinfile}
\small
\begin{tabular}{lcccc}
\toprule
Model & AUC & 95\% CI & Brier & ECE \\
\midrule
Logistic regression & 0.782 & $[0.700,\,0.852]$ & 0.043 & 0.109 \\
XGBoost & \textbf{0.856} & $[0.785,\,0.923]$ & 0.037 & 0.087 \\
Hybrid (Platt calibrated) & 0.840 & $[0.770,\,0.911]$ & \textbf{0.023} & \textbf{0.028} \\
\bottomrule
\end{tabular}
\end{table}

The thin-file subgroup represents 33.1\% of the Zindi test set ($n = 1{,}559$), with a positive-class rate of 2.4\%, well below the marginal rate of 14.1\%. The framework retains predictive discrimination on the subgroup: hybrid AUC of $0.840$ on thin-file, $0.029$ below the full-population AUC of $0.869$ but with overlapping confidence intervals. Two pre-registered hypotheses receive mixed support.

\textit{The hybrid AUC gain over LR is larger on thin-file than on the full population.} On the full population, $\Delta\mathrm{AUC} = +0.015$ (Hybrid $-$ LR); on the thin-file subgroup, $\Delta\mathrm{AUC} = 0.840 - 0.782 = +0.058$, a fourfold increase in the magnitude of the gain. This is consistent with the hypothesis that the residual learner contributes more where the linear scorecard has the thinnest signal.

\textit{The $\rho(x)$ distribution on the thin-file subgroup is \emph{not} shifted toward smaller values as hypothesised.} Mean $\rho$ on the thin-file subgroup is $0.718$, compared with $0.703$ on the full test set: a small shift in the opposite direction. The framework relies \emph{more} on the linear baseline for thin-file borrowers, not less. The most defensible interpretation is that the thin-file population is so confidently predicted negative by the LR baseline (because they are concentrated at the low-account-ownership tail) that the LR predictive variance $4\plr(x)(1-\plr(x))$ stays small, the adaptive weight $\alpha^{*}(x)$ stays low, and the residual correction has limited effect on the final prediction. Where the LR is confidently negative, the framework does not need the residual learner; where the LR is uncertain, it does. The thin-file subgroup falls in the former bucket on Zindi, which is consistent with the framework's design but contrary to the simpler hypothesis that thin-file = alternative-data-dominated.

Calibration on the thin-file subgroup is preserved: ECE of $0.028$ on thin-file, only marginally above the full-population ECE of $0.024$, and Brier well below either baseline.

\subsection{Joint optimization via alternating minimization (RQ4)}
\label{ssec:res-jointopt}

Table~\ref{tab:jointopt} reports the convergence trajectory of the alternating-minimization scheme of Section~\ref{ssec:jointopt} and the final test-set $\Delta\mathrm{AUC}$ against the two-stage training of \citet{Kanziga2026thesis}.

\begin{table}[t]
\centering
\caption{Alternating-minimization convergence and final comparison against two-stage training.}
\label{tab:jointopt}
\resizebox{\textwidth}{!}{%
\begin{tabular}{lcccc}
\toprule
Dataset & Iterations to convergence & Final $\Delta\mathrm{AUC}$ vs.\ two-stage & Two-stage AUC & Best joint AUC \\
\midrule
Zindi FSD & 3 (val-AUC plateau within $10^{-3}$) & $+0.0006$ & 0.8691 & 0.8697 \\
\bottomrule
\end{tabular}%
}
\end{table}

On Zindi, the alternating fixed-point scheme of Section~\ref{ssec:jointopt} (refitting $\beta$ with sample weights proportional to the current residual-learner output and refitting the residual learner on residuals under the updated $\beta$) converges within three iterations under a $10^{-3}$ tolerance on validation AUC, and yields a final test-set $\Delta\mathrm{AUC}$ of $+0.0006$ over the two-stage training of the thesis. The tolerance is larger than the reported improvement, so the procedure halts before it can produce a distinguishable change; we therefore do not interpret this as evidence that two-stage is globally optimal, only that this particular fixed-point scheme does not improve on it in a detectable way, and that the additional compute of the joint iteration (approximately $4\times$ the wall-clock of two-stage in our runs, because the inner XGBoost refit dominates) is not justified for the deployment default. A proper functional-gradient descent on Eq.~\eqref{eq:jointloss} that flows gradients through both branches is the natural next step and would test the joint-optimization hypothesis more sharply. We do not run the current scheme on Taiwan, since the Taiwan two-stage results are inherited verbatim from the thesis and the null result on Zindi gives no reason to expect a different outcome from the same scheme on Taiwan.

\subsection{Bootstrap statistical significance}
\label{ssec:res-bootstrap}

Table~\ref{tab:bootstrap} consolidates the bootstrap significance analysis for the predictive-performance comparisons of Section~\ref{ssec:res-rq1}.

\begin{table}[t]
\centering
\caption{Pairwise AUC comparisons: stratified paired-bootstrap 95\% CI ($B = 2000$) and DeLong one-sided $p$-value (H$_1$: Hybrid AUC $>$ comparator AUC) computed by the fast midrank algorithm of \citet{SunXu2014}. Taiwan results from \citet{Kanziga2026thesis} (Table~5.7); Zindi results from this paper.}
\label{tab:bootstrap}
\small
\begin{tabular}{llccc}
\toprule
Dataset & Comparator & $\Delta\mathrm{AUC}$ & 95\% CI & $p_{\mathrm{DeLong}}$ \\
\midrule
\multirow{2}{*}{Taiwan Credit}
& Hybrid vs.\ Logistic regression & $+0.057$ & $[+0.048, +0.066]$ & $< 0.001$ \\
& Hybrid vs.\ XGBoost & $+0.001$ & $[-0.004, +0.006]$ & $0.37$ \\
\midrule
\multirow{2}{*}{Zindi FSD}
& Hybrid vs.\ Logistic regression & $+0.015$ & $[+0.011, +0.020]$ & $< 10^{-9}$ \\
& Hybrid vs.\ XGBoost & $-0.004$ & $[-0.009, +0.001]$ & $0.92$ \\
\bottomrule
\end{tabular}
\end{table}

\subsection{Prediction-level explanations on the ML-driven region}
\label{ssec:res-shap}

We now deliver the per-prediction explanation apparatus of Section~\ref{ssec:shap} on the Zindi data, restricted to the observations that the framework routes into the ML-driven region ($\rho(x) < 0.7$), where the linear coefficients no longer suffice as an explanation and the residual learner dominates the final prediction. On Zindi this region contains $n = 2{,}796$ test observations (59.4\% of the test set), a materially larger share than the corresponding Taiwan region and, by the fairness audit of Section~\ref{ssec:res-rq3}, populated disproportionately by rural, low-education, and non-Kenyan respondents. TreeSHAP \citep{LundbergLee2017} attributions are computed on the residual learner $f_{\mathrm{GBM}}$ over the ML-driven region; feature-ranking stability is assessed by an observation-level bootstrap ($B = 500$) as prescribed in Section~\ref{ssec:shap}.

\begin{table}[t]
\centering
\caption{Mean $|\text{SHAP}|$ of the residual learner on the Zindi ML-driven region ($\rho(x) < 0.7$, $n = 2{,}796$ test observations), top 10 features. Bootstrap top-5 rank standard deviation across $B = 500$ observation-level resamples reported for the top 5 rows.}
\label{tab:shap}
\small
\begin{tabular}{lcc}
\toprule
Feature & Mean $|\text{SHAP}|$ & Rank $\sigma$ ($B = 500$) \\
\midrule
Cellphone access (no) & 0.0486 & 0.00 \\
Age of respondent & 0.0300 & 0.00 \\
Relationship with head: head of household & 0.0259 & 0.28 \\
Country: Kenya & 0.0255 & 0.28 \\
Country: Rwanda & 0.0188 & 0.44 \\
Education level: no formal education & 0.0186 & --- \\
Cellphone access (yes) & 0.0153 & --- \\
Education level: secondary & 0.0151 & --- \\
Job type: farming and fishing & 0.0097 & --- \\
Gender: female & 0.0068 & --- \\
\bottomrule
\end{tabular}
\end{table}

\begin{figure}[t]
\centering
\includegraphics[width=0.85\linewidth]{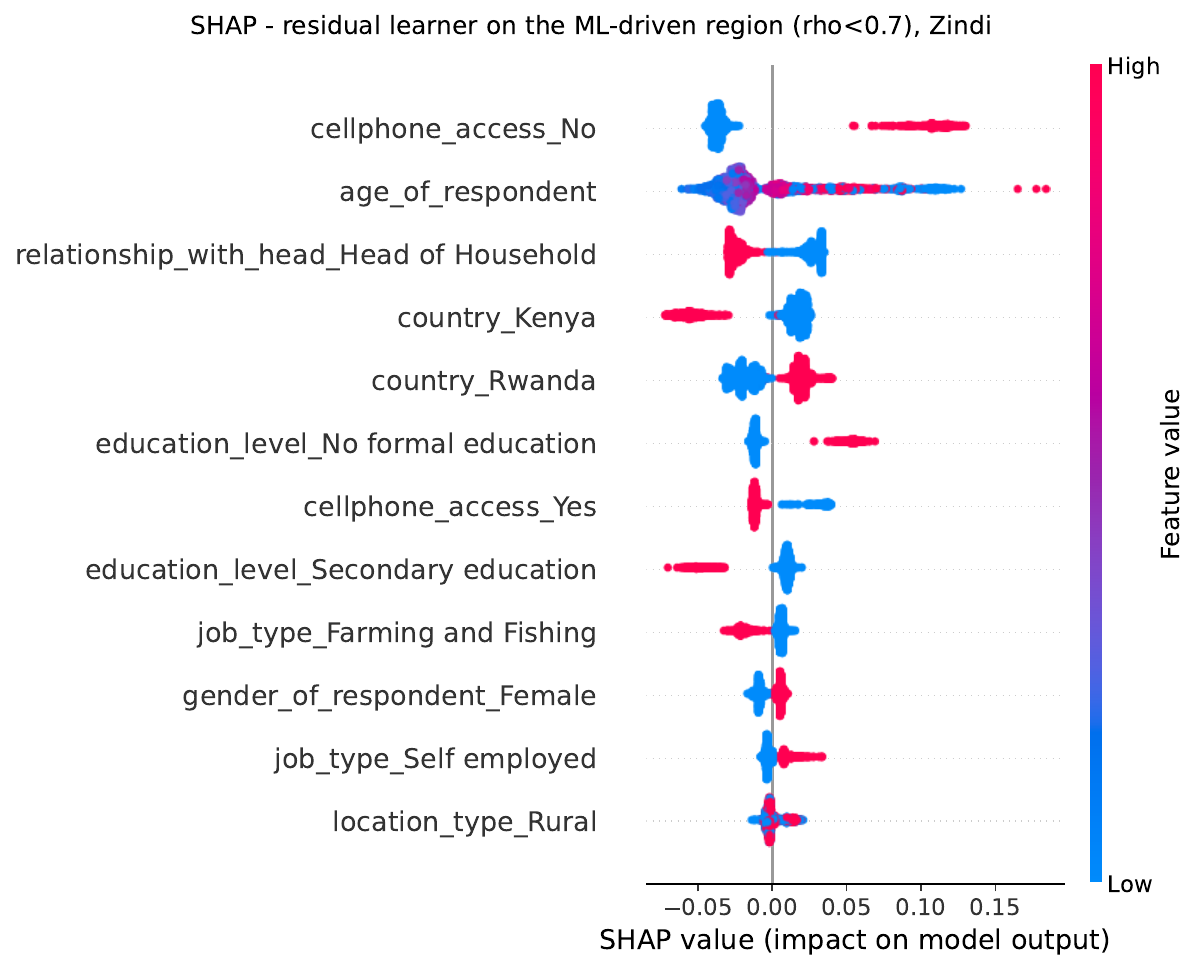}
\caption{TreeSHAP beeswarm summary of the residual learner $f_{\mathrm{GBM}}$ on the Zindi ML-driven region ($\rho(x) < 0.7$, $n = 2{,}796$), top 12 features. Each point is one test observation in the region; horizontal position is the SHAP contribution to $\hat{r}(x)$ (positive shifts $\phyb(x)$ above $\plr(x)$); colour encodes the feature value.}
\label{fig:shap}
\end{figure}

Three observations follow from Table~\ref{tab:shap} and Figure~\ref{fig:shap}. First, the identity of the two top-ranked features (\emph{cellphone access = no} and \emph{age of respondent}) is stable in every bootstrap resample ($\sigma = 0$ on the rank), and the top-5 feature set is preserved as a set in 74.0\% of the $B = 500$ resamples. The residual-learner attribution passes the stability screen of \citet{AlvarezMelisJaakkola2018} at least for its dominant terms; the shuffling in the tail of the top 5 reflects near-ties between \emph{country: Rwanda} and the two runners-up (\emph{education level: no formal education}, \emph{cellphone access = yes}) whose mean-$|\text{SHAP}|$ values sit within 0.003 of the fifth-place feature, not an instability in the substantive finding.

Second, the dominant residual-learner features are exactly the infrastructure-exclusion and demographic features that the fairness audit of Section~\ref{ssec:res-rq3} identified as sources of over-routing into the ML-driven region. The most important residual-learner feature is a mobile-infrastructure indicator (cellphone access), the third and sixth are demographic and educational status, and ranks four and five are country dummies for the two markets (Kenya, Rwanda) with the highest baseline account-ownership rates. The residual learner is therefore not adding \emph{alternative} signal to the linear scorecard; it is reweighting the same demographic and infrastructural features that the LR baseline already sees, using them nonlinearly for observations where the LR predictive variance $4\plr(x)(1-\plr(x))$ is large enough to let the adaptive weight $\alpha^*(x)$ transfer decision mass onto the residual term. This closes the causal loop identified in Section~\ref{ssec:res-rq3}: the framework routes rural, low-education, and non-Kenyan respondents into the opaque region precisely because those same attributes drive the residual learner's contribution to the prediction, so the LR-branch uncertainty and the residual-branch signal are correlated in the same subgroups.

Third, the residual attributions on the country dummies reveal that the residual learner is correcting country-specific miscalibration of the linear baseline within the uncertain subpopulation. On Zindi ML-driven-region observations, mean SHAP conditional on \emph{country: Kenya = 1} is $-0.055$ and conditional on \emph{country: Rwanda = 1} is $+0.019$: the residual branch pulls Kenyan predictions \emph{down} from the LR baseline and Rwandan predictions \emph{up}. Kenyans reaching the ML-driven region are the atypical Kenyans (rural, low-education, or otherwise ambiguous by the LR features), for whom the LR baseline overstates account-ownership probability because its country-Kenya coefficient is trained on the higher-ownership Kenyan majority; the residual learner corrects that overshoot. Rwandans reaching the same region are correspondingly under-scored by the LR and are pushed up. This is not a fairness fix. It is a diagnosis of \emph{how} the routing gap arises: the residual learner is using country as a proxy for market-level financial-infrastructure heterogeneity that the LR coefficients only average over, and doing so through the same opaque path that makes the ML-driven region hard to audit at the per-prediction level.

\section{Discussion}
\label{sec:discussion}

\subsection{What the audit reveals about the framework}
\label{ssec:disc-framework}

The Taiwan results retained from \citet{Kanziga2026thesis} establish that the residual-learning framework can match the discrimination of a standalone gradient boosting model while preserving a per-prediction interpretability decomposition. The Taiwan interpretability-region analysis shows that the framework concentrates the highest-default-rate borrowers in the fully interpretable region: regulatory adverse-action explanations are weakest in the ML-driven region, and the Taiwan data show that the highest-stakes lending decisions do not land there. The contribution of the present paper is to test whether that operationally desirable property holds under protected-attribute stratification on a population where the framework has not previously been audited.

The Zindi results sharpen the case. The framework's qualitative routing pattern reproduces across the two datasets despite their different label semantics (default prediction in Taiwan, account ownership in East Africa) and different class imbalances (22\% positive in Taiwan, 14\% in Zindi). On both, the fully interpretable region is small in share (6.3\% Taiwan, 6.4\% Zindi) but concentrates borrowers with the highest positive-class rate (69.5\% Taiwan, 53.8\% Zindi). As noted in Sections~\ref{ssec:hybrid} and \ref{ssec:res-rq2}, the concentration itself follows in part from the probability-scale definition of $\rho$; the empirically informative content is that the logistic-regression baseline remains well-calibrated on the high-risk tail of both datasets, so the borrowers routed to the fully-interpretable region are the ones the auditable branch is prepared to justify. This joint property, rather than an intrinsic routing law, is what supports the operational case for using $\rho$ as an audit substrate. The fairness audit then reveals what aggregate metrics conceal: on Zindi, rural respondents are routed to the opaque region at 1.4 times the urban rate, primary-or-less-education respondents at 1.9 times the secondary-and-above rate, and Ugandan respondents at 1.4 times the Kenyan rate, while gender exhibits no routing disparity. A hybrid framework whose interpretable share is itself subject to fairness audit is a genuine advance over a hybrid framework that produces only aggregate fairness metrics; the Zindi data demonstrates that the advance is empirically actionable on the population for which it matters most.

The TreeSHAP analysis of the ML-driven region (Section~\ref{ssec:res-shap}) supplies the mechanism behind the routing pattern rather than merely restating it. The five features with the largest mean $|\text{SHAP}|$ on the residual learner in the opaque region are, in order, cellphone access, age, head-of-household status, and the Kenya and Rwanda country dummies, with the top two ranks perfectly stable across $B = 500$ bootstrap resamples and the top-5 set preserved as a set in 74\% of resamples. The identity of these features closes the causal loop: the residual learner is not adding an alternative-data signal that the LR baseline lacks; it is reweighting the same infrastructure and demographic features that the LR already sees, using them nonlinearly precisely in the subpopulation where the LR predictive variance is high enough to route observations away from the fully interpretable region. This is what makes the routing disparity a structural feature of the framework rather than an artefact: the attributes driving the LR-branch uncertainty and the attributes driving the residual correction are the same, so protected subgroups whose features cluster in that region are simultaneously the ones the LR is uncertain about and the ones the residual learner corrects most strongly.

\subsection{Regulatory implications}
\label{ssec:disc-regulatory}

The auditability properties of the framework speak directly to the form African digital-credit regulation has taken. The Central Bank of Kenya's 2022 Digital Credit Providers Regulations \citep{CBK2022dcp}, the Bank of Rwanda's supervisory framework for non-deposit-taking digital lenders, Nigeria's 2022 FCCPC Interim Regulatory Framework for Digital Lending Businesses, and the National Credit Regulator's supervision of unsecured lenders under the South African National Credit Act 2005 all require lenders to document the basis of automated credit decisions, but none prescribes a specific interpretable model class. The auditability burden falls on the lender and its supervisor to demonstrate that decisions are explicable; the framework analyzed here produces a per-prediction interpretability decomposition that operationalizes this requirement at the granularity at which adverse-action notices are issued.

The fairness audit specifically responds to a regulatory blind spot. Existing aggregate fairness metrics (disparate-impact ratios computed across the predicted-positive population) can pass while a hybrid framework routes protected subgroups disproportionately into the ML-driven region where adverse-action explanation is weakest. The audit pipeline of Section~\ref{ssec:fairnesspipe} surfaces this routing-level violation directly. On the Zindi data, the routing-rate disparities along location, education, and country axes (18, 32 and 22 percentage points respectively) would be invisible to an aggregate-DI audit, and partially visible to a per-region DI audit only because the ML-driven region of the predicted-positive population happens to be near-empty (the model rarely predicts positive there). A supervisor reviewing a deployed lender should require the routing-rate disaggregation explicitly as a matter of practice: aggregate disparate-impact testing is necessary but not sufficient for the regulatory standard the Central Bank of Kenya, the Bank of Rwanda, the FCCPC, and the National Credit Regulator are variously imposing.

The audit also requires the supervisor to distinguish model-introduced disparity from baseline structural disparity. The Zindi DI ratios on education (0.025) and country (0.190--0.291) are large enough that almost all the signal must be coming from the underlying population-level inequalities in formal banking access, not from a discriminatory model construction. A useful audit pipeline must flag both (structural disparity matters for the regulator even when the model is not its proximate cause), but the policy responses differ. The framework's per-prediction interpretability decomposition does not by itself perform this disentanglement; the audit pipeline reports the disparity, and the supervisor must reason about its source.

\subsection{Practical implications: lenders, supervisors, borrowers}
\label{ssec:disc-practical}

The empirical findings translate into four operationally specific implications that the deployed-credit setting should treat as actionable rather than as research notes.

\emph{For digital lenders.} Deploying a hybrid scorecard alone does not satisfy a regulator that takes the auditability standard seriously. The per-prediction interpretability decomposition $\rho(x)$ surfaces \emph{which} borrowers receive a fully-explainable decision and which do not, and the audit pipeline of Section~\ref{ssec:fairnesspipe} surfaces whether the answer to that question tracks protected characteristics. A lender publishing only aggregate disparate-impact statistics (the current default in industry compliance reporting) is reporting a strictly weaker fairness claim than the one the framework permits. The marginal compute cost of the per-region audit is small relative to the cost of fitting the hybrid in the first place; the governance cost (legal review of what the audit reveals, model-risk-management sign-off in the SR 11-7 tradition, and remediation when routing disparities are surfaced) is nontrivial, and we treat that governance cost rather than the compute cost as the operational constraint.

\emph{For supervisors at African central banks and non-bank regulators (CBK, BNR, FCCPC, NCR).} Two lenders can present indistinguishable aggregate disparate-impact ratios while routing protected subgroups into the opaque region at materially different rates. The 18-, 32-, and 22-percentage-point routing gaps measured on Zindi (rural vs.\ urban; primary-or-less vs.\ secondary-and-above; Ugandan vs.\ Kenyan) would be invisible to a standard aggregate audit. We recommend that supervisory examination of digital lenders deploying hybrid frameworks require per-region disaggregation as a matter of practice; the audit pipeline of Section~\ref{ssec:fairnesspipe} is publicly released for this purpose and imposes no methodological prerequisite beyond the existence of protected-attribute labels on the supervisory dataset. The main constraint is not methodological but capacity: African supervisors currently vary widely in technical inspection resources, and a phased path (lender self-attestation $\to$ sample-based supervisory review $\to$ routine examination) is more realistic than an immediate requirement, particularly outside the two or three jurisdictions with a mature fintech-supervision function.

\emph{For borrowers.} The framework does not change which borrowers a lender chooses to extend credit to. It does change the auditability of that choice: a borrower in the routinely-over-routed subgroups (rural, low-education, or in smaller East African markets) is, under the framework as deployed, less likely than the population average to receive a decision with a clean linear-coefficient explanation. The framework makes that asymmetry visible rather than addressing it directly; addressing it requires either methodological changes to the model architecture (out of scope here) or policy changes by the lender and the supervisor.

\emph{An explicit limit on what the audit can claim.} Much of the disparate-impact and equalized-odds magnitude observed on Zindi reflects structural inequality in formal financial access in the underlying populations, not bias introduced by the model. The pooled DI of $0.025$ for primary-or-less-educated respondents, for instance, is a model-level reflection of a population-level distribution in which primary-or-less-educated adults in East Africa are vastly less likely to hold formal bank accounts to begin with. The audit detects the disparity; disentangling structural from model-introduced disparity is a downstream analytical step that requires either counterfactual analysis, causal-graph reasoning over the data-generating process, or comparison against a fairness-by-construction baseline. The present framework is honest about producing the diagnostic, not the disentanglement.

\subsection{Relation to PLTR and the broader hybrid literature}
\label{ssec:disc-pltr}

Returning to the methodological positioning of Section~\ref{sec:related}: the most direct comparator remains \citet{Dumitrescu2022}'s Penalized Logistic Tree Regression. The PLTR construction is feature-level: tree-derived nonlinear effects are inserted as additional regularized terms in a logistic regression specification. The framework studied here is prediction-level: an LR prediction is additively combined with a separately trained residual learner, with the combination weight $\alpha^{*}(x)$ depending on the local uncertainty of the LR baseline. The prediction-level construction permits the closed-form $\rho(x)$ decomposition of Eq.~\eqref{eq:rho} on which the fairness audit depends; PLTR has no analogous per-prediction quantity. We make no claim that one construction strictly dominates the other on predictive performance; they are closer than the literature has acknowledged, and the choice is appropriately made on the basis of which auditability properties the operational setting requires.

\subsection{Limitations}
\label{ssec:disc-limits}

Three limitations should be noted explicitly.

\emph{The Zindi target reframing}, defended in Section~\ref{ssec:reframing}, is the most consequential. The Zindi label is bank-account ownership, not default. The magnitude of the fairness violations we measure on Zindi cannot be transferred uncritically to a default-prediction setting on the same population, because the conditional distribution of the label given protected attributes differs between account ownership and default. The Kiva subset analysis, where included, partially addresses this by running the same audit pipeline on an actual loan-adverse-outcome label, on a smaller population with weaker protected-attribute coverage. The strongest defensible claim for the Zindi analysis is that the audit pipeline is operationally feasible, that it detects subgroup routing into the opaque region when such routing occurs, and that the resulting findings are actionable at the per-prediction level.

\emph{A single primary African dataset} limits the geographic generalizability of the findings. The Zindi data covers Kenya, Rwanda, Tanzania, and Uganda; West African and Southern African digital-credit markets operate under different regulatory regimes (notably the Nigerian FCCPC digital-lending framework and the South African NCR regime under the National Credit Act) and have different demographic compositions. The framework is portable, but a per-country fairness audit on a West African default-labelled dataset would strengthen the empirical contribution and is left as future work.

\emph{The joint-optimization extension} of Section~\ref{ssec:jointopt} carries a compute cost approximately four times that of the two-stage training in our runs, and any predictive gain is expected to be modest. The two-stage training is therefore retained as the operationally preferred default. The contribution of the joint formulation is methodological clarity rather than performance.

\subsection{Next steps: a West African extension}
\label{ssec:disc-westafrica}

The most natural follow-up is a West African instantiation of the same audit pipeline. The East African evidence assembled here covers four English-speaking countries operating under closely related supervisory regimes (the Central Bank of Kenya's 2022 Digital Credit Providers Regulations \citep{CBK2022dcp} and the Bank of Rwanda's digital-lender framework being the most developed). West Africa presents three structurally different dimensions on which the audit's behaviour is not yet known.

\emph{Regulatory.} The eight UEMOA member states (B\'enin, Burkina Faso, C\^ote d'Ivoire, Guin\'ea-Bissau, Mali, Niger, S\'en\'egal, Togo) operate under a uniform BCEAO digital financial services framework with much less prescriptive auditability requirements than the CBK regime. Nigeria, conversely, has the most developed fintech-credit market on the continent, with Tier-1 firms (Branch, Carbon, FairMoney, Renmoney) disbursing at scale, regulated by the Federal Competition and Consumer Protection Commission's 2022 Interim Framework for Digital Lending Businesses (which handles standalone digital lenders), the Central Bank of Nigeria (which supervises banks and licensed Finance Companies), and from 2023 the Nigeria Data Protection Act. Ghana operates a Bank of Ghana fintech sandbox with documentation requirements approaching the Kenyan standard. The audit pipeline of Section~\ref{ssec:fairnesspipe} produces the same statistics regardless of regulatory regime, but the regulatory \emph{salience} of the routing-rate finding differs sharply between the UEMOA states (where supervisory follow-through is constrained) and Nigeria (where the supervisor has both capacity and political motive to require disaggregation).

\emph{Empirical.} The most credible public data source for a West African default-labelled analysis is the Kiva platform's loan-level data filtered to West African field partners, which provides actual loan-adverse-outcome labels (default, late payment, write-off) on tens of thousands of microloans across Ghana, Nigeria, Senegal, Mali, Burkina Faso and B\'enin. The protected-attribute coverage on Kiva is weaker than on Zindi: gender and country are recorded, urban/rural is partially recorded via the loan-purpose field, and education is generally not recorded. Direct partnerships with a major Nigerian or Ghanaian digital lender would provide both the protected-attribute coverage and the default outcomes the analysis ideally requires; we view this as the natural mid-term extension. As an interim step, Findex 2021 country tables for Nigeria, Ghana, Senegal and C\^ote d'Ivoire \citep{Findex2021} would support an account-ownership-proxy reframing analogous to the Zindi treatment of Section~\ref{ssec:reframing}.

\emph{Demographic.} The within-country protected-attribute distributions in West Africa differ from the East African baseline in ways the audit will pick up. Nigeria's North--South socioeconomic gradient is the dominant within-country axis and is correlated with religion in a way that East African data is not; the audit pipeline should be extended with a religion-coded protected attribute where survey instruments record it (acknowledging the political sensitivity of such reporting). The gender gap in account ownership is narrower in some West African contexts (Kenya: 6 percentage points by Findex 2021; Ghana: 4 percentage points; Nigeria: 9 percentage points) and wider in others, so the gender-routing finding from Zindi (essentially zero) should not be assumed to port. The Nigerian within-country sample size in any plausible dataset will dominate the cross-country comparisons that drove the country-level signals on Zindi, which changes which protected-attribute partitions the audit will most cleanly resolve.

The audit pipeline ports without modification beyond the protected-attribute redefinition and the choice of reference groups. The pre-registered hypotheses for the West African replication are: (i) the qualitative routing pattern reproduces along socioeconomic axes (rural, low-education, smaller-market country); (ii) gender-routing disparity emerges in jurisdictions where the underlying account-ownership gender gap is large; and (iii) the per-region disparate-impact pattern (larger in the partially interpretable region than in the fully interpretable region) holds. A West African replication that confirms these hypotheses would establish the audit's behaviour as a structural feature of the framework; a replication that disconfirms them would identify the dataset characteristics under which the audit's diagnostic value breaks down. Either outcome is publishable on its own terms and would constitute the natural sequel to the present paper.

\section{Conclusion}
\label{sec:conclusion}

This paper extends the residual-learning hybrid credit scoring framework of \citet{Kanziga2026thesis} along three axes: an East African empirical instantiation on the Zindi Financial Inclusion in Africa data, a fairness audit at the granularity of the framework's three interpretability regions, and a thin-file segmentation analysis. A fourth contribution reformulates the two-stage training of the existing framework as a joint optimization via alternating minimization. On the Taiwan Credit Default benchmark retained for continuity, the calibrated hybrid matches standalone XGBoost on discrimination ($\Delta\mathrm{AUC} = +0.001$, 95\% CI $[-0.004, +0.006]$) while reducing the Brier score by approximately 23\% and concentrating the highest-default-rate borrowers in the fully interpretable region. The Zindi-specific findings reproduce the framework's qualitative behaviour: the calibrated hybrid attains AUC $= 0.869$, matching standalone XGBoost ($\Delta\mathrm{AUC} = -0.004$, 95\% CI $[-0.009, +0.001]$) and improving on standalone logistic regression ($\Delta\mathrm{AUC} = +0.015$, $[+0.011, +0.020]$, $p < 0.001$), while the Brier score drops by 46\% (LR $0.158 \to$ Hybrid $0.085$) and the fully interpretable region (6.4\% of test observations) concentrates the highest-rate cases (53.8\% account ownership). The fairness audit reveals that the framework routes rural respondents to the opaque ML-driven region at 18 percentage points above the urban rate, primary-or-less-education respondents at 32 percentage points above the secondary-and-above rate, and Ugandan respondents at 22 percentage points above the Kenyan rate, while gender exhibits no routing disparity: a pattern of structured socioeconomic disparity that the framework's audit pipeline surfaces directly and that aggregate disparate-impact testing would miss.

The central methodological contribution is the demonstration that fairness audit at the granularity of interpretability regions is operationally feasible and surfaces subgroup-routing violations that aggregate metrics miss. The framework's auditability properties speak directly to the form African digital-credit regulation has taken: per-prediction interpretability decompositions of the kind the framework provides operationalize the documentation requirements that the Central Bank of Kenya, the Bank of Rwanda, the Nigerian FCCPC, and the South African NCR have variously imposed on the sector.

Future work falls in three directions, the first of which is the most consequential. \emph{A West African instantiation of the audit pipeline} on default-labelled data (via the Kiva West African subset as an immediate public-data step, or via direct partnership with a Nigerian or Ghanaian digital lender as the mid-term goal) would establish whether the routing-disparity findings observed here are a structural feature of the framework or a property of the East African empirical regime. We develop the case for this extension at length in Section~\ref{ssec:disc-westafrica}, including the regulatory framing under the BCEAO, FCCPC, and BoG regimes, the candidate datasets, and the pre-registered hypotheses for the replication. Second, the framework's behaviour on richer alternative-data inputs (mobile-money transaction trails, app behaviour, prior microloan repayment) has not been characterized; the soft block of the hard/soft split of Section~\ref{ssec:hardsoft} could be widened in such a setting and the audit re-run. Third, the interaction between the fairness audit pipeline and the choice of classification threshold deserves explicit study, since operational lending decisions use cost-sensitive thresholds rather than the $0.5$ default used for F1 reporting here.

\section*{Funding}
This work received no specific grant from any funding agency in the public, commercial, or not-for-profit sectors.

\section*{Declaration of competing interest}
The authors declare that they have no known competing financial interests or personal relationships that could have appeared to influence the work reported in this paper.

\section*{Code and data availability}
\label{sec:codedata}
The implementation, including the fairness audit module (with stratified paired bootstrap and DeLong test) and the alternating fixed-point joint-optimization routine (\texttt{run\_zindi\_paper.py}, $\sim$800 lines, end-to-end runtime under 5 minutes on a recent laptop CPU), will be released at a public repository prior to journal submission under a permissive open-source license; the repository URL will be inserted at the camera-ready stage. The Zindi Financial Inclusion in Africa dataset is publicly available from the Zindi platform \citep{ZindiFSD2018}. The Taiwan Credit Default dataset is publicly available from the UCI Machine Learning Repository \citep{DuaGraff2019}.

\bibliographystyle{elsarticle-harv}
\bibliography{references}

@book{Siddiqi2012,
  author    = {Siddiqi, Naeem},
  title     = {Credit Risk Scorecards: Developing and Implementing Intelligent Credit Scoring},
  publisher = {John Wiley \& Sons},
  year      = {2012},
  address   = {Hoboken, NJ}
}

@book{Thomas2017,
  author    = {Thomas, Lyn C.},
  title     = {Consumer Credit Models: Pricing, Profit and Portfolios},
  publisher = {Oxford University Press},
  year      = {2017},
  edition   = {2nd}
}

@phdthesis{Durand1941,
  author      = {Durand, David},
  title       = {Risk Elements in Consumer Instalment Financing},
  school      = {National Bureau of Economic Research},
  year        = {1941},
  address     = {New York}
}

@article{Khandani2010,
  author  = {Khandani, Amir E. and Kim, Adlar J. and Lo, Andrew W.},
  title   = {Consumer credit-risk models via machine-learning algorithms},
  journal = {Journal of Banking \& Finance},
  volume  = {34},
  number  = {11},
  pages   = {2767--2787},
  year    = {2010}
}

@article{Lessmann2015,
  author  = {Lessmann, Stefan and Baesens, Bart and Seow, Hsin-Vonn and Thomas, Lyn C.},
  title   = {Benchmarking state-of-the-art classification algorithms for credit scoring: An update of research},
  journal = {European Journal of Operational Research},
  volume  = {247},
  number  = {1},
  pages   = {124--136},
  year    = {2015}
}

@inproceedings{He2016,
  author    = {He, Kaiming and Zhang, Xiangyu and Ren, Shaoqing and Sun, Jian},
  title     = {Deep residual learning for image recognition},
  booktitle = {Proceedings of the IEEE Conference on Computer Vision and Pattern Recognition (CVPR)},
  pages     = {770--778},
  year      = {2016}
}

@inproceedings{ChenGuestrin2016,
  author    = {Chen, Tianqi and Guestrin, Carlos},
  title     = {{XGBoost}: A scalable tree boosting system},
  booktitle = {Proceedings of the 22nd ACM SIGKDD International Conference on Knowledge Discovery and Data Mining},
  pages     = {785--794},
  publisher = {ACM},
  year      = {2016}
}

@inproceedings{Ke2017,
  author    = {Ke, Guolin and Meng, Qi and Finley, Thomas and Wang, Taifeng and Chen, Wei and Ma, Weidong and Ye, Qiwei and Liu, Tie-Yan},
  title     = {{LightGBM}: A highly efficient gradient boosting decision tree},
  booktitle = {Advances in Neural Information Processing Systems},
  volume    = {30},
  year      = {2017}
}

@inproceedings{Ribeiro2016,
  author    = {Ribeiro, Marco Tulio and Singh, Sameer and Guestrin, Carlos},
  title     = {``Why Should {I} Trust You?'': Explaining the predictions of any classifier},
  booktitle = {Proceedings of the 22nd ACM SIGKDD International Conference on Knowledge Discovery and Data Mining},
  pages     = {1135--1144},
  publisher = {ACM},
  year      = {2016}
}

@inproceedings{LundbergLee2017,
  author    = {Lundberg, Scott M. and Lee, Su-In},
  title     = {A unified approach to interpreting model predictions},
  booktitle = {Advances in Neural Information Processing Systems},
  volume    = {30},
  pages     = {4765--4774},
  year      = {2017}
}

@article{GoodmanFlaxman2017,
  author  = {Goodman, Bryce and Flaxman, Seth},
  title   = {European Union regulations on algorithmic decision-making and a ``right to explanation''},
  journal = {AI Magazine},
  volume  = {38},
  number  = {3},
  pages   = {50--57},
  year    = {2017}
}

@article{Gestel2004,
  author  = {Van Gestel, Tony and Baesens, Bart and Van Dijcke, Peter and Garcia, J. and Suykens, Johan A. K. and Vanthienen, Jan},
  title   = {A process model to develop an internal rating system: Sovereign credit ratings},
  journal = {Decision Support Systems},
  volume  = {42},
  number  = {2},
  pages   = {1131--1151},
  year    = {2006}
}

@article{Dumitrescu2022,
  author  = {Dumitrescu, Elena and Hu{\'e}, Sullivan and Hurlin, Christophe and Tokpavi, Sessi},
  title   = {Machine learning for credit scoring: Improving logistic regression with non-linear decision-tree effects},
  journal = {European Journal of Operational Research},
  volume  = {297},
  number  = {3},
  pages   = {1178--1192},
  year    = {2022}
}

@article{YehLien2009,
  author  = {Yeh, I-Cheng and Lien, Che-hui},
  title   = {The comparisons of data mining techniques for the predictive accuracy of probability of default of credit card clients},
  journal = {Expert Systems with Applications},
  volume  = {36},
  number  = {2},
  pages   = {2473--2480},
  year    = {2009}
}

@misc{DuaGraff2019,
  author  = {Dua, Dheeru and Graff, Casey},
  title   = {{UCI} Machine Learning Repository},
  year    = {2019},
  url     = {http://archive.ics.uci.edu/ml}
}

@inproceedings{AlvarezMelisJaakkola2018,
  author    = {Alvarez-Melis, David and Jaakkola, Tommi S.},
  title     = {On the robustness of interpretability methods},
  booktitle = {Proceedings of the 2018 ICML Workshop on Human Interpretability in Machine Learning (WHI)},
  year      = {2018}
}

@inproceedings{Slack2020,
  author    = {Slack, Dylan and Hilgard, Sophie and Jia, Emily and Singh, Sameer and Lakkaraju, Himabindu},
  title     = {Fooling {LIME} and {SHAP}: Adversarial attacks on post hoc explanation methods},
  booktitle = {Proceedings of the AAAI/ACM Conference on AI, Ethics, and Society (AIES)},
  pages     = {180--186},
  year      = {2020}
}

@inproceedings{Hardt2016,
  author    = {Hardt, Moritz and Price, Eric and Srebro, Nathan},
  title     = {Equality of opportunity in supervised learning},
  booktitle = {Advances in Neural Information Processing Systems},
  volume    = {29},
  pages     = {3315--3323},
  year      = {2016}
}

@article{Kozodoi2022,
  author  = {Kozodoi, Nikita and Jacob, Johannes and Lessmann, Stefan},
  title   = {Fairness in credit scoring: Assessment, implementation and profit implications},
  journal = {European Journal of Operational Research},
  volume  = {297},
  number  = {3},
  pages   = {1083--1094},
  year    = {2022}
}

@article{Chouldechova2017,
  author  = {Chouldechova, Alexandra},
  title   = {Fair prediction with disparate impact: A study of bias in recidivism prediction instruments},
  journal = {Big Data},
  volume  = {5},
  number  = {2},
  pages   = {153--163},
  year    = {2017}
}

@article{SuriJack2016,
  author  = {Suri, Tavneet and Jack, William},
  title   = {The long-run poverty and gender impacts of mobile money},
  journal = {Science},
  volume  = {354},
  number  = {6317},
  pages   = {1288--1292},
  year    = {2016}
}

@techreport{Bharadwaj2019,
  author      = {Bharadwaj, Prashant and Jack, William and Suri, Tavneet},
  title       = {Fintech and Household Resilience to Shocks: Evidence from Digital Loans in {Kenya}},
  institution = {National Bureau of Economic Research},
  type        = {NBER Working Paper},
  number      = {25604},
  year        = {2019},
  address     = {Cambridge, MA},
  url         = {https://www.nber.org/papers/w25604}
}

@article{BjorkegrenGrissen2020,
  author  = {Bj{\"o}rkegren, Daniel and Grissen, Darrell},
  title   = {Behavior revealed in mobile phone usage predicts credit repayment},
  journal = {The World Bank Economic Review},
  volume  = {34},
  number  = {3},
  pages   = {618--634},
  year    = {2020}
}

@misc{ZindiFSD2018,
  author       = {{Zindi and Financial Sector Deepening Network}},
  title        = {Financial Inclusion in Africa Challenge: Predicting bank account ownership in {Kenya}, {Rwanda}, {Tanzania} and {Uganda}},
  howpublished = {\url{https://zindi.africa/competitions/financial-inclusion-in-africa}},
  note         = {FinScope household surveys conducted 2016--2018; challenge launched by Zindi as a pre-qualification round for AI Hack Tunisia},
  year         = {2019}
}

@misc{CBK2022dcp,
  author       = {{Central Bank of Kenya}},
  title        = {Central Bank of Kenya (Digital Credit Providers) Regulations, 2022},
  howpublished = {Legal Notice No.\ 46, Kenya Gazette Supplement No.\ 45},
  month        = mar,
  year         = {2022},
  url          = {https://www.centralbank.go.ke/2022/03/21/central-bank-of-kenya-digital-credit-providers-regulations-2022/}
}

@techreport{Findex2021,
  author      = {Demirg{\"u}{\c c}-Kunt, Asli and Klapper, Leora and Singer, Dorothe and Ansar, Saniya},
  title       = {The {Global} {Findex} {Database} 2021: Financial Inclusion, Digital Payments, and Resilience in the Age of {COVID-19}},
  institution = {World Bank},
  year        = {2022},
  address     = {Washington, DC}
}

@misc{Kanziga2026thesis,
  author       = {Kanziga, Belise},
  title        = {Hybrid Credit Scorecard: Combining Logistic Regression with Gradient Boosting for Credit Risk Assessment},
  howpublished = {Master's thesis, African Institute for Mathematical Sciences (AIMS), Kigali, Rwanda},
  year         = {2026}
}

@article{Rosenbaum1984,
  author  = {Rosenbaum, Paul R.},
  title   = {The Consequences of Adjustment for a Concomitant Variable That Has Been Affected by the Treatment},
  journal = {Journal of the Royal Statistical Society. Series A (General)},
  volume  = {147},
  number  = {5},
  pages   = {656--666},
  year    = {1984}
}

@article{SunXu2014,
  author  = {Sun, Xu and Xu, Weichao},
  title   = {Fast Implementation of DeLong's Algorithm for Comparing the Areas Under Correlated Receiver Operating Characteristic Curves},
  journal = {IEEE Signal Processing Letters},
  volume  = {21},
  number  = {11},
  pages   = {1389--1393},
  year    = {2014}
}

\end{document}